\documentclass[twocolumn,10pt]{IEEEtran}
\usepackage{graphicx}
\usepackage{framed}
\usepackage{epstopdf}
\usepackage{amsmath}
\usepackage{amsthm}
\usepackage{amsfonts}
\usepackage[justification=centering]{caption}
\usepackage{soul} % µŒÈë soul °ü
\usepackage{color, xcolor}
\theoremstyle{}
\newtheorem{theorem}{Theorem}
\newtheorem{lemma}{Lemma}
\newtheorem{definition}{Definition}
\newtheorem{corollary}{Corollary}
\newtheorem{proposition}{Proposition}
\newtheorem{delivery}{Delivery Strategy}

\newtheorem{example}{Example}
\newtheorem{remark}{Remark}

\usepackage{bm}
\usepackage{booktabs}
\usepackage{multirow}
\usepackage{algorithm}
\usepackage{algpseudocode}
\usepackage{subfigure}
\usepackage{amssymb}
\usepackage{cite}
\usepackage{multicol}
\usepackage{lipsum}% dummy text
\usepackage{pstricks}

\newcommand{\tabcaption}{\def\@captype{table}\caption}
\newcommand{\tabincell}[2]{\begin{tabular}{@{}#1@{}}#2\end{tabular}}
\begin{document}
\title{{Asymptotically Optimal Coded Distributed Computing via Combinatorial Designs}\\
\author{\IEEEauthorblockN{Minquan Cheng,  Youlong Wu, Xianxian Li, and Dianhua Wu}
\thanks{M. Cheng and X. Li are with Guangxi Key Lab of Multi-source Information Mining $\&$ Security, Guangxi Normal University,
Guilin 541004, China  (e-mail: chengqinshi@hotmail.com, lixx@gxnu.edu.cn).}
\thanks{Y. Wu is with the School of Information Science and Technology, ShanghaiTech University, Shanghai 201210, China  (e-mail:  wuyl1@shanghaitech.edu.cn).}
\thanks{D. Wu is with the School of Mathematics and Statistics, Guangxi Normal University, Guilin 541006, China (e-mail:  dhwu@gxnu.edu.cn).}
}
}

\maketitle

\begin{abstract}
Coded distributed computing (CDC) introduced by Li \emph{et al.} can greatly reduce the communication load for MapReduce computing systems. In the cascaded CDC with $K$ workers, $N$ input files and $Q$ output functions,  each input file will be mapped by $r$ workers and each output function will be computed by $s$ workers such that coding techniques can be applied to create multicast opportunities. The main drawback of most existing CDC  schemes is that they require the original data to be split into a large number of input files that grows exponentially with $K$, which would significantly increase the coding complexity and degrade the system performance.
In this paper, we first use a classical combinatorial structure $t$-design, for any integer $t\geq 2$, to develop a low-complexity and communication-efficient CDC with $r=s$. Our scheme has much smaller  $N$ and  $Q$ than the existing schemes under the same parameters $K$, $r$, and $s$; and achieves smaller communication loads compared with the state-of-the-art schemes when $K$ is relatively large. Remarkably, unlike the previous schemes that realize on large operation fields, our scheme operates in one-shot communication on the minimum binary field $\mathbb{F}_2$. With a derived lower bound on the communication load under one-shot linear delivery, we show that the $t$-design scheme is asymptotically optimal.  Furthermore, we show that our construction method can incorporate the other combinatorial structures that have a similar property to $t$-design. For instance, we use $t$-GDD to obtain another one-shot asymptotically optimal CDC scheme over $\mathbb{F}_2$ that has different parameters from $t$-design.  Finally, we show that our construction method can also be used to construct CDC schemes with $r\neq s$ that have small file number and output function number.
\end{abstract}


%
%\begin{abstract}
%xxxx
%\end{abstract}
%
%\begin{IEEEkeywords}
%xxxx
%\end{IEEEkeywords}

\section{Introduction}
 Distributed computing systems have been widely applied to execute large-scale computing tasks, since they can greatly speed up task execution by letting distributed computing workers execute computation jobs in parallel and exploiting distributed computing and storage resources. However, when exchanging a massive amount of data among the computing workers, distributed computing systems would suffer a severe communication bottleneck due to the limited communication resource and high transmitted traffic load.  For instance,  in the TeraSort \cite{Guo2017iShuffleIH} and SelfJoin \cite{SelfJoin}  applications running in Amazon EC 2 cluster,   the time cost of data exchange occupies $65\%\sim 70\%$ of the overall job execution time \cite{CZMJS}.

Coded distributed computing (CDC), proposed by Li \emph{et al.} in \cite{LMYA}, is considered as a promising approach to reduce the communication load for distributed computing systems such as MapReduce \cite{DG} and Spark \cite{ZCFSS}, by introducing repetitive computations on the input data to create coding multicast opportunities. Consider a cascaded $(K,r,s,N,Q)$ MapReduce system consisting of $K$ workers, $N$ input files and $Q$ output functions where each input file will be mapped by $r$ workers and each output function will be calculated by $s$ workers. Here $r$ represents the computation load\footnote{In \cite{LMYA}, the computation load is defined as the total number of computed Map functions at the nodes, normalized by $N$. This paper focuses on the symmetric case where each file will be   mapped by exactly $r$ workers.}  and  $s\geq 1$ allows to support multi-round MapReduce tasks where the output values in the previous round serve as the inputs to the Map functions of the next round.  The MapReduce system contains ``Map", ``Shuffle" and ``Reduce" phases. In the Map phase, each worker stores and maps the local input files, obtaining $Q$ intermediate values of equal size $T$-bit for each input file. In the Shuffle phase, the workers generate coded symbols from the local IVs, and multicast them to other workers such that all desired intermediate values can be recovered by the desired workers. In the Reduce phase, each output value will be reduced by the assigned $s$ workers based on the locally computed and recovered intermediate values. Let communication load be the total communication bits normalized by $NQT$. In \cite{LMYA}, it shows that the proposed CDC scheme can achieve the minimum communication load under a specific output function assignment.
%When $s=1$, i.e., each output value is computed exactly once, the  CDC scheme in \cite{LMYA} is similar to the coded caching scheme for the D2D network \cite{Ji2014D2D}.  }

The CDC approach has attracted wide attention, and many works have been conducted focusing on the scalability and optimality of CDC in various settings.  For instance,   the linear dependency of intermediate values and the properties of map functions were exploited to improve the computation-communication trade-off \cite{horii2020improved}. Saurav \emph{et al.} described the MapReduce computations on graphs and leveraged the graph structure to create coded multicast opportunities \cite{compresscoded}. The resource allocation problem of CDC has been investigated in \cite{CDC1,chen2020coded}, in which optimal CDC schemes to minimize the total execution time were proposed. The heterogeneous CDCs with different storages and computation loads among workers were considered in \cite{woolsey2021new,xu2021new,wang2021batch,reisizadeh2019coded,kim2019optimal,Wang22}. For more works on CDCs, please see the survey in \cite{CDCsurvey}.

\subsection{ Research Motivation}
When $s=1$ and $Q=K$, i.e., each output function is computed once and each worker computes one output function, the CDC problem is equivalent to the coded caching problem for device-to-device (D2D) networks  \cite{Ji2014D2D}. In this paper, we mainly focus on the cascaded case $s> 1$, in which the coded caching schemes for D2D networks are not applicable anymore.
 In  \cite{JQ}, the authors generated the cascaded CDC scheme by using the schemes generated by placement delivery array (PDA) in \cite{YCTC} $s$ rounds.   It is worth noting that the PDA was originally proposed to reduce the subpacketization of coded caching problem when each worker requests distinct content, not able to characterize the worker demands if some workers request the same content. This leads to the  PDA-based  CDC schemes \cite{JQ} failing to exploit the common intermediate values desired by multiple workers, and thus incurring redundant communication load.

  % There are many results on the coded caching schemes for D2D network, such as \cite{YKSC,WCYT}. It is worth noting that the authors in \cite{WCYT} point out that we can use the coded caching scheme generalized by placement delivery array (PDA) in \cite{YCTC} for the shared-link network to obtain the scheme for D2D network. As a result all the results of PDAs in \cite{YCTC,YTCC,SZG,CJYT,CJTY,CJWY,CWZW,CLZW,WCWC,ZCW,MW,ZCJ,ZWCC,Agrawal'ISIT, Li2022PlacementDA}%in \cite{YCTC,YTCC,CJTY,CJWY,CJYT,MN,SZG,CJTY} can be used to design the coded caching schemes for D2D network.

%For the cascaded case where each output function is computed $s> 1$ times, the schemes for the D2D network can not be directly used to generate the cascaded CDC schemes. The authors in \cite{JQ} generated the cascaded CDC scheme by using the schemes generated by PDA as in \cite{WCYT} for the D2D network $s$ times. These schemes have an operation field $\mathbb{F}_2$ and are one-shot deliveries.

In order to reduce the redundant communication load,
%For the cascaded case where each output function is computed $s\geq 1$ times,
most existing CDC schemes adopt some combinatorial structures to design the data placement and output function assignment  by means of the Maximum separable (MDS) codes or the randomly linear combination method.  For instance, the first and classical cascaded CDC scheme \cite{LMYA} uses all the $r$-subset and $s$-subset of $[K]$ to design the data placement and output function assignment respectively where $1\leq r,s\leq K$. Then in the Shuffle phase, for any subset $\mathcal{S}\subseteq [K]$ with a cardinality $|\mathcal{S}|\in\{\max\{r+1,s\},\ldots, \min\{r+s,K\}\}$, each worker $k\in\mathcal{S}$ will use MDS codes to generate and deliver linear combinations of intermediate value segments based on its locally computed intermediate values  such that each linear combination is decodable by the exclusively desired workers in $\mathcal{S}\backslash\{k\}$, i.e., achieving a multicast gain $|\mathcal{S}|-1$. The authors in \cite{LMYA} showed that their scheme achieves the minimum communication load under this specific output function assignment, but at the cost of requiring exponentially large numbers of both input files and output functions in terms of $\binom{K}{r}$ and $\binom{K}{s}$, respectively. This leads to unexpected performance losses in practical implementations when  $K$ is relatively large. For example, in the CodedTeraSort experiment sorting $12$ GB data with $K = 16$ workers, $r= 6$, and $100$ Mbps network speed, each worker needs to generate all file indices and initialize $\binom{K}{r+1}$ multicast groups to transfer all intermediate values to the intended workers. The corresponding time cost will dominate the overall execution time\cite{LMYA}.

To reduce the required numbers of both input files and output functions, a hypercube scheme where the data placement and output function assignment are generated by the hypercube structure was proposed in \cite{WCJ} for the case $r=s$. In their scheme, the communications take place in multiple rounds with multicast gains equal to  $r-1,r,\ldots, \min\{2r-1,K-1\}$, in each of which one worker generates linear combinations based on the local intermediate values and broadcasts them to the other workers. The authors also showed that when $r=s=2$ their scheme can achieve a strictly smaller communication load than that  the scheme in \cite{LMYA}. However, their scheme still requires exponentially large numbers of both input files and output functions in terms of $(\frac{K}{r})^{r-1}$ and achieves a larger communication load when $r=s>2$ compared with the scheme in \cite{LMYA}.  The authors in \cite{JWZ} used a symmetric balanced incomplete block design (SBIBD) to generate the data placement and output function assignment to obtain a new CDC scheme with $K=N=Q$. In their scheme, each coded symbol carries contents desired by $r-1$ workers, i.e., the multicast gain is $r-1$.
%The authors \cite{JQ} used the placement delivery array (PDA) to generate cascaded CDC schemes. These schemes have an operation field $\mathbb{F}_2$ and are one-shot deliveries.  It is worth noting that the PDA was originally proposed to reduce the subpacketization of coded caching problem when each worker requests distinct contents, not able to characterize the worker demands if some workers request the same content \cite{YCTC,YTCC,SZG,CJYT,CJTY,CJWY,CWZW,CLZW,WCWC,ZCW,MW,ZCJ,ZWCC }. This leads to the  PDA-based  CDC schemes \cite{JQ} failing to exploit the common intermediate values desired by multiple workers, and thus incurring redundant communication load.
We list all the above results in Table \ref{tab-known-CDC}.
{\begin{table*}
  \centering
  \renewcommand\arraystretch{1}
    \setlength{\tabcolsep}{1mm}{
  \caption{Existing cascaded CDC schemes\label{tab-known-CDC} }
  \begin{tabular}{|c|c|c|c|c|c|c|c|}
\hline
Parameters &\tabincell{c}{Worker number\\  $K$} & \tabincell{c}{Computation \\  Load $r$}& \tabincell{c}{Replication \\ Factor $s$}&  \tabincell{c}{Number of \\ Files $N$}   & \tabincell{c}{Number of output \\ Functions $Q$}&  \tabincell{c}{Communication \\ Load $L_{\text{comm.}}$}& \tabincell{c}{Operation \\ Field $\mathbb{F}_2$}\\
\hline
\tabincell{c}{\cite{JQ}: $K,r,s\in \mathbb{N}^{+}$,\\ $1\leq r,s\leq K$} &
$K$& $r$ &$s$ &$(\frac{K}{r})^{r-1}$& $\frac{K}{\text{gcd}(K,s)}$&$\frac{s}{r-1}(1-
\frac{r}{K})$ &Yes\\
\hline
\tabincell{c}{\cite{LMYA}: $K,r,s\in \mathbb{N}^{+}$,\\ $1\leq r,s\leq K$}    & $K$  & $r$ & $s$ & ${K\choose r}$ & ${K\choose s}$& \tabincell{c}{$\sum\limits_{l=\max\{r+1,s\}}^{\min\{r+s,K\}}$\\ $\left( \frac{l-r}{l-1}\cdot\frac{{K-r\choose K-l}{r\choose l-s}}{{K\choose s}}\right)$}&No\\
 \hline

%\tabincell{c}{\cite{WCJ}: $K,r\in \mathbb{N}^{+}$, \\ $K$  is divisible by $r$} &$K$&$r$&$r$& $(\frac{K}{r})^{r-1}$ &$(\frac{K}{r})^{r-1}$&{\color{red}\tabincell{c}{$\frac{r^r(K-r)}{K^r(r-1)}+\sum\limits_{l=2}^r\left[ (\frac{r}{K})^{r+l}\right.$\\ $\left.{r\choose l}{K/r\choose 2}^l \frac{2^l l}{2l-1}\right]$}}&No\\
%\hline

 \tabincell{c}{\cite{WCJ}: $K,r\in \mathbb{N}^{+}$, \\ $K$  is divisible by $r$} &$K$& $r$&$r$& $(\frac{K}{r})^{r}$ &$(\frac{K}{r})^{r}$&
\tabincell{c}{$\frac{1}{2}-\frac{1}{2}\cdot \left(\frac{r}{K}\right)^{r} $\\
$+\left(1-\frac{r}{K}\right)^r \cdot \frac{1}{4r-2}$}&No\\
\hline

\tabincell{c}{\cite{JWZ}: $(K,r,\lambda)$ SBIBD, \\ $K,r,\lambda\in \mathbb{N}^{+}$,\\ $2\leq \lambda\leq r$ } &
$K$& $r$ &$r$ &$K$& $K$&$\frac{r}{r-1}\cdot\frac{K-r}{K}$ &No\\
\hline

\end{tabular}
}
\end{table*}}

\subsection{Contribution and Organization}
In this paper, we focus on constructing the new cascaded case with  $s=r$. Different from the constructing method of the existing schemes in \cite{LMYA,JWZ,JQ,WCJ}, we first let $N=Q$ and use the same strategy to design the data placement and output function assignment. Then we find that if the data placement satisfies the cross property (which is introduced before Definition \ref{def-design}) then we can design a delivery strategy over $\mathbb{F}_2$ such that the multicast gain is as large as possible, i.e., the communication load is as small as possible. With the above idea, we propose novel schemes based on combinatorial structures to reduce the communication load and the required numbers of input files and output functions. Our contributions are summarized as follows.

{\begin{table*}
  \centering
  \renewcommand\arraystretch{1}
    \setlength{\tabcolsep}{1mm}{
  \caption{New schemes where $t$, $N$, $M$, $\lambda \in \mathbb{N}^{+}$ and $2\leq t\leq M\leq N$.
  \label{tab-Main-results}}
  \begin{tabular}{|c|c|c|c|c|c|c|c|}

\hline
Parameters &\tabincell{c}{Worker number\\  $K$} & \tabincell{c}{Computation \\  Load $r$}& \tabincell{c}{Replication \\ Factor $s$}&  \tabincell{c}{Number of \\ Files $N$}   & \tabincell{c}{Number of output \\ Functions $Q$}&  \tabincell{c}{Communication \\ Load $L_{\text{comm.}}$}& \tabincell{c}{Operation \\ Field $\mathbb{F}_2$}\\
\hline

Theorem \ref{th-design-CDC}&
$\frac{\lambda{N\choose t}}{{M\choose t}}$ & $\frac{KM}{N}$ & $\frac{KM}{N}$ & $N$ & $N$& $\frac{N-1}{2N}<\frac{1}{2}$&Yes\\ [4pt]
\hline

Theorem \ref{th-GDD-CDC}&$\frac{\lambda{m\choose t}q^t}{{M\choose t}}$& $\frac{KM}{mq}$&$\frac{KM}{mq}$& $mq$ &$mq$&$\frac{1}{2}+\frac{q-2}{2mq}$&Yes\\
\hline

Theorem \ref{th-design-CDC-rs}& $\frac{\lambda{N\choose t}}{{M\choose t}}$ & $\frac{KM}{N}$  &$\frac{\lambda(N-t+1)}{ M-t+1}$ &$N$& ${N\choose t-1}$&$\frac{N-t+1}{Nt}<\frac{1}{t}$&Yes \\
\hline
\end{tabular}
}
\end{table*}}

\begin{itemize}
\item  Based on classical combinatorial structure $t$-design \cite{CD} where $t\geq 2$, we obtain our first scheme, i.e., the scheme for Theorem \ref{th-design-CDC} in Table \ref{tab-Main-results}. For the special $2$ design, our scheme achieves a multicast gain $r+s-2=2r-2$, which is only one less than the maximum multicast gain of any $(K,r,s=r,N,Q)$ scheme under one-shot linear delivery. To relax the parameter restriction in the $t$-design scheme, we propose a scheme based on $t$-group divisible design ($t$-GDD), i.e., the scheme for Theorem \ref{th-GDD-CDC} in Table \ref{th-GDD-CDC}. We further show that our constructing method can also be extended to the case $r\neq s$,  i.e., the scheme for Theorem \ref{th-design-CDC-rs} in Table \ref{th-GDD-CDC}.%,%The $t$-GDD scheme is feasible for more flexible parameters.
%Finally, we show that our constructing method can also be extended to the case $r\neq s$, which is the scheme for Theorem \ref{th-design-CDC-rs} in Table \ref{tab-Main-results}. By comparison with the existing schemes {\red in \cite{LYMA,WCJ,JQ}}, our third scheme has a significant advantage in terms of the number of files.

%It is worth noting that by relaxing the cross property, our construction method also allows us to obtain other low-complexity and communication-efficient CDC schemes. For example, we apply $t$-GDD (which is a generation of $t$-design) \cite{CD} to obtain another one-shot asymptotically optimal CDC scheme.
% and has different parameters from the $t$-design scheme for Theorem \ref{th-design-CDC} (see Remark \ref{remark-relation} in Subsection \ref{sec-desgin}). The $t$-GDD scheme for Theorem \ref{th-GDD-CDC} is listed in Table \ref{tab-Main-results}.

\item Compared to the state-of-the-art schemes, both the $t$-design  and $t$-GDD schemes can reduce the values of $N$ and $Q$, while achieving less communication load. Comparing the $t$-design  and $t$-GDD schemes, the latter allows more flexible system parameters, at the cost of slightly increasing the communication load. %also further reduces the communication load when we use the special $2$-design, i.e., BIBD; our $t$-GDD scheme also has significant advantages in reducing the values of $N$ and $Q$ while achieving less communication load.

\item We derive a lower bound on the minimum communication load under one-shot linear delivery, and prove that both $t$-design and $t$-GDD schemes are asymptotical optimal under one-shot linear delivery when the number of workers is sufficiently large.

\end{itemize}

The rest of this paper is organized as follows. Section \ref{sec-pre} describes
the system model. Section \ref{sec-combin} introduces some concepts   of combinatorial structures that will be useful to our scheme design.  Section \ref{sec-main} introduces our main results including two new schemes and their performance analyses. Section \ref{Sec:Schemes} provides the detailed construction of the two proposed cascaded CDC schemes. Section \ref{sec-extension} provides an extension scheme for the case $r\neq s$. Finally, we conclude the paper in Section \ref{sec-conclusion}.

\subsection{Notations} In this paper,  we use the following notations unless otherwise stated.
\begin{itemize}
\item Bold capital letters, bold lowercase letters, and calligraphic fonts will be used to denote arrays, vectors, and sets, respectively.
\item We assume that all the sets are in increasing order; for a set $\mathcal{V}$, we let $\mathcal{V}(j)$ represent the $j$-th smallest element of $\mathcal{V}$ and let $\mathcal{V}(\mathcal{J})=\{\mathcal{V}(j)| j\in\mathcal{J} \}$.
\item   $|\cdot|$ is used to represent the cardinality of a set or
the length of a vector.
\item For any positive integers $a$, $b$, $t$ with $a<b$ and $t\leq b $, and any nonnegative set $\mathcal{V}$,
 let $[a,b]=\{a,a+1,\ldots,b\}$, especially $[1,b]$ be shorten by $[b]$, and
${[b]\choose t}=\{\mathcal{V}\ |\   \mathcal{V}\subseteq [b], |\mathcal{V}|=t\}$, i.e., ${[b]\choose t}$ is the collection of all $t$-sized subsets of $[b]$. We use $a|b$ to denote that $b$ is divisible by $a$.
\end{itemize}

\section{Problem Formualtion}
\label{sec-pre}
\begin{figure}[h]
	\centering
	\includegraphics[scale=0.28]{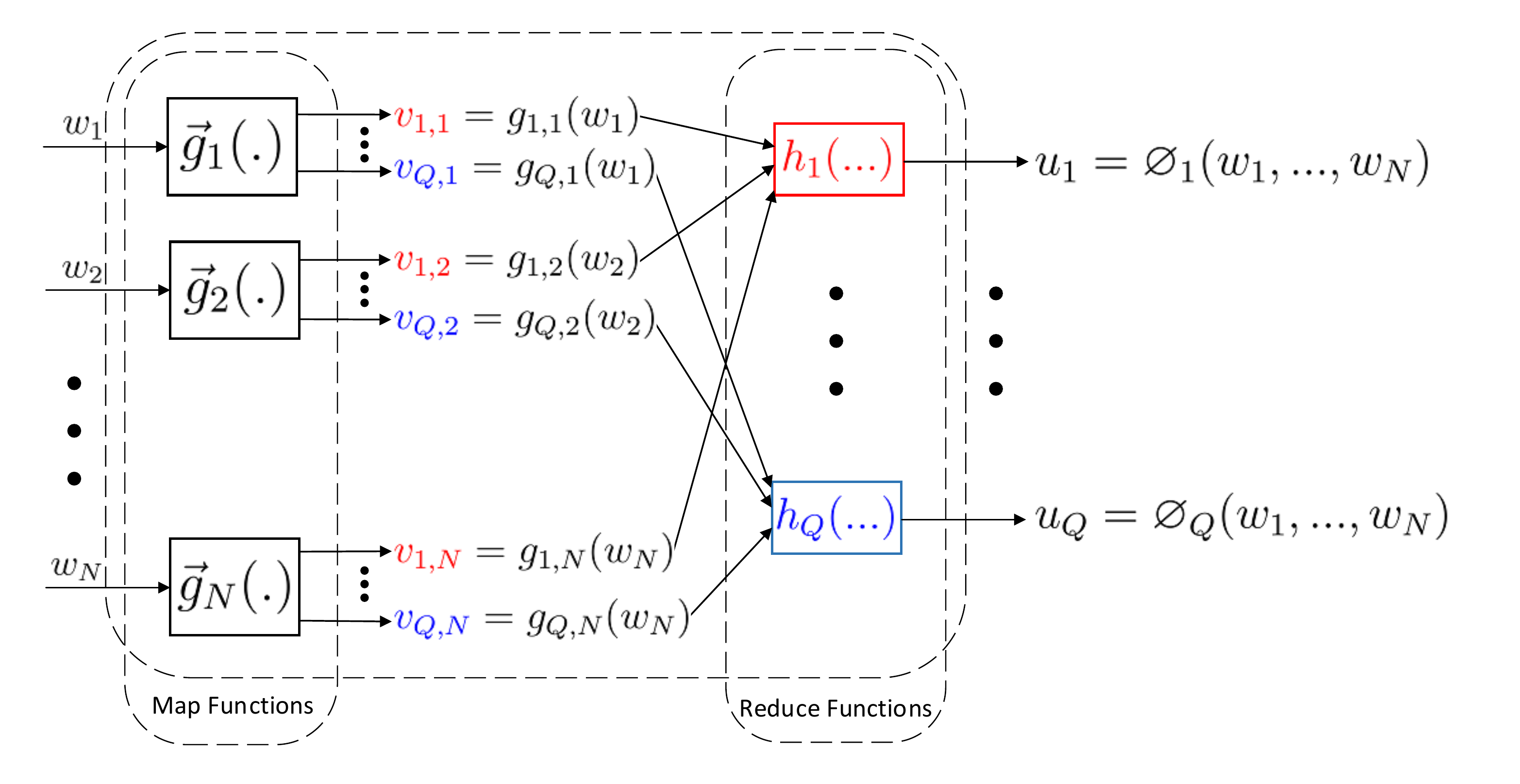}
	\caption{Illustration of a two-stage distributed computing framework. The
overall computation is decomposed into computing a set of Map and Reduce
functions.}
	\label{fig:model}
\end{figure}

%In this section, we formulate the cascaded coded distributed computing problem and
%\subsection{Cascaded Coded Distributed Computing System}
\label{secSystem}
Consider a $(K,r,s,N,Q)$ coded distributed computing system. There
are $K$ distributed computing workers who will compute $Q$ output functions from $N$ input files each of equal size. Denote the worker set, $N$ files, and $Q$ output functions by $\mathcal{K}$, $\mathcal{W}=\{w_1,w_2,\ldots,w_N\}$, and $\mathcal{Q}=\{\phi_1,\phi_2,\ldots,\phi_Q\}$, respectively.   For each function $\phi_q$ where $q\in [Q]$, the output function will be computed by some workers  as
\begin{align*}
u_q &\triangleq\phi_q(w_{1},w_{2},\ldots,w_{N})\\
&\triangleq h_q(g_{q,1}(w_{1}),g_{q,2}(w_{2}),\ldots,g_{q,N}(w_{N})) \in \mathbb{F}_{2^E},
 \end{align*}for some integer $E$ where $u_q$ is the output computed from output function $\phi_q$,  $g_{q,n}(\cdot)$ is  called Map function, and $h_q$ is called Reduce function, for $q\in [Q]$ and $n\in [N]$. The parameter $v_{q,n}\triangleq g_{q,n}(w_{n}) \in \mathbb{F}_{2^T}$ where $q\in [Q]$ and $n\in [N]$ is called intermediate value.   We assume that
each output function is assumed to be computed by $s\in[K]$ workers and each input file will be  mapped by $r$ workers.
%For each function $\phi_q$ where $q\in [Q]$, let $\mathcal{A}_q\in {[K]\choose s}$ represent the worker set each of which is arranged to compute the output function
% We assume that each intermediate value has $T$ bits, for some positive $T$. In order to support multiple-round computing where the reduced results of the previous round are the inputs of the next round Map operation, each output function is assumed to be computed by $s\in[K]$ workers. {\green We assume that each input file will be  mapped by $r$ workers.}
 As illustrated in Fig. \ref{fig:model}, a $(K,r,s,N,Q)$  cascaded coded distributed computing scheme consists of the following three phases.
\begin{itemize}
\item {\bf Map phase.}
%Each worker $k\in \mathcal{K}$ first stores $M$ files, denoted by $\mathcal{Z}_k$.
Let $\mathcal{D}_n\in {[K]\choose r}$, for $n\in [N]$, represent the worker set each of which stores file $w_n$. Then the files stored by worker $k\in  [{K}]$ can be written as follows.
\begin{align}\label{eq-caches}
\mathcal{Z}_k=\{w_{n}\ |\ k\in \mathcal{D}_n, n\in [N]\}.
\end{align} Using the stored files in \eqref{eq-caches} and Map functions $\{g_{q,n}(\cdot)\}$ where $q\in [Q]$ and $n\in [N]$, worker $k$ could compute the following intermediate values
\begin{align*}
\mathcal{I}_k=\{v_{q,n}=g_{q,n}(w_{n})\ |\ q\in [Q],k\in \mathcal{D}_n,n\in [N]\}.
\end{align*} Similar to the existing works \cite{LMYA,JWZ,JQ,WCJ}, we also assume that each worker maps the same number of files, say $M$ files, i.e., $|\mathcal{Z}_k|=M$ for each $k\in [K]$. Since each file is mapped by $r\in[K]$ workers,  we have $Nr=KM$.

\item {\bf Shuffle phase.} Let $\mathcal{A}_q\in {[K]\choose s}$, for  $q\in [Q]$,  represent the worker set, each of which is arranged to compute the output function $\phi_q$. Then each worker $k\in [K]$ is arranged to compute the output functions as follows:
\begin{align}
\label{eq-computing-task}
\mathcal{Q}_k=\{u_q=h_q(v_{q,1},v_{q,2}, \ldots,v_{q,N})\ |\ k\in \mathcal{A}_q, q\in [Q]\}.
\end{align}Since each worker can not store all the files, all the workers should exchange their locally computed intermediate values to ensure that each worker can compute the output function. Assume that worker $k\in \mathcal{K}$ sends a coded message $X_k$
with length $l_k$
generated by its locally computed intermediate values to the other workers.

\item {\bf Reduce phase.} Based on the received messages $\{X_1,X_2,\ldots,X_K\}$ and its locally computed intermediate values in $\mathcal{I}_k$, worker $k\in  [K]$ can compute each output function in $\mathcal{Q}_k$.
\end{itemize}

%Let  $|v_{q,n}|$ denote the bits of intermediate value $v_{q,n}$.
Define the \emph{multicast gain}  as the total  bits of intermediate values desired by all workers divided by the  total number of transmitted bits, i.e.,
\begin{align} \label{eq-mul-gain}
g=\frac{\sum\limits_{k=1}^K \sum\limits_{q\in\mathcal{Q}_k}\sum\limits_{n\in [N]\backslash\mathcal{Z}_k} |v_{q,n}|}{\sum\limits_{k=1}^K|X_k|} =\frac{sQ(N-M)T}{\sum\limits_{k=1}^K l_k}.
\end{align} Following the same definition in \cite{LMYA},  we  define the communication load as the total number of communication bits in the Shuffle phase normalized by $QNT$, i.e.,
%\begin{align}\label{eq-computing-load}
%r\triangleq\frac{\sum_{k=1}^{K}|\mathcal{Z}_k|}{N}= \frac{KM}{N},
%\end{align}and
\begin{align}\label{eq-communication-load}
L_{\text{comm.}}\triangleq\frac{\sum_{k=1}^{K}|X_k|}{QNT}=\frac{\sum_{k=1}^{K}l_k}{QNT}=\frac{s(1-M/N)}{g}.
\end{align}%That is,
%$r$ is the average number of workers that map each file and
 %$L_{\text{comm.}}$ is the ratio of the amount of transmitted data to $QNT$.
We refer to a scheme as \emph{one-shot linear} communication if each transmitted signal is a linear combination of intermediate values and can be directly decoded by the desired workers in one-round communication with a linear operation. Denote $g^*_\text{one-shot}$ as the maximum multicast gain achieved by the one-shot linear schemes.

From \eqref{eq-caches} and \eqref{eq-computing-task}, the data placement and  output function arranged strategies are determined by $\{\mathcal{D}_n\}$  and $\{\mathcal{A}_q\}$. This implies that the key step of designing a CDC step is to design file stored set $\mathfrak{D}$ and the output function arranged set $ \mathfrak{A}$  defined as
\begin{align}
\mathfrak{A}\triangleq\{\mathcal{A}_1,\mathcal{A}_2,\ldots,\mathcal{A}_Q\}, \ \
\mathfrak{D}\triangleq\{\mathcal{D}_1,\mathcal{D}_2,\ldots,\mathcal{D}_N\}.
\label{eq-families}
\end{align}

Given the system parameters $(K,r,s,N,Q)$,  we aim to design the output function arranged set $\mathfrak{A}$, the file stored set $\mathfrak{D}$, and the corresponding delivery strategy such that the communication load is minimized.

By letting $\mathfrak{D}={[K]\choose r}$ and $ \mathfrak{A}={[K]\choose s}$, the authors in \cite{LMYA} proposed the first well-known CDC scheme achieving the following communication load.
\begin{lemma}\rm
\label{lemma-LMYA}
For any positive integers $K$, $r$ and $s$ there exists a  $(K,r,s,N={K\choose r}, Q={K\choose s})$ CDC scheme with transmission load
 \begin{align}\label{eq-converse}
L_{\text{LMYA}}(r,s)=\sum\limits_{l=\max\{r+1,s\}}^{\min\{r+s,K\}}\left(\frac{{K-r\choose K-l}{r\choose l-s}}{{K\choose s}}\cdot \frac{l-r}{l-1}\right).
\end{align}
\end{lemma}
%\subsection{Previous Works}
%\begin{remark}\rm\label{LMYAResults}\blue
%\begin{itemize}
%\item

\begin{remark}\rm 
The authors in \cite{LMYA} proved that $L_{\text{LMYA}}(r,s)$ in \eqref{eq-converse} is optimal under a  specific output function assignment satisfying $ \mathfrak{A}={[K]\choose s}$,   for any file stored set design. However this scheme involves exponentially large file number $N$ and output function number $Q$, i.e.,  $N/\binom{K}{r}\in\mathbb{N}$ and $Q/\binom{K}{s}\in\mathbb{N}$, which limits the application scenarios and may lead to unexpected performance losses
in practical implementations \cite{JWZ,LMYA,JQ,WCJ}.
\end{remark}

There are some works focusing on reducing the values of $N$ and $Q$ such as \cite{JWZ,JQ,WCJ}. It is very interesting that when $r=s=2$ under a delicate design $\mathfrak{A}$, a lower communication load than \eqref{eq-converse} is achievable in \cite{JWZ}. Unfortunately, it is not known whether there exists any scheme breaking the optimality when  $r=s>2$.

%\begin{remark}\rm
%There exist some coded caching schemes \cite{Agrawal'ISIT, Li2022PlacementDA} addressing subpacketization problems based on combinatorial designs. However, their schemes focus on the case where each user requires one file and each file is requested by a single user, i.e., $|\mathfrak{A}|=K$ and each element of $\mathfrak{A}$ is a $1$-subset, not suitable for the cascaded CDC where each element of $\mathfrak{A}$ is a {\red subset of size $s\geq 2$.}
%\end{remark}

In this paper, we focus on constructing low-complexity and communication-efficient schemes that operate over $\mathbb{F}_2$ and reduce the communication load while keeping small values of $N$ and $Q$.

\section{Preliminary: Combinatorial Design Structures}
\label{sec-combin}
  In this section, we introduce some concepts of combinatorial design that will be useful to construct our schemes.

\begin{definition}[\cite{CD}, Design]\rm
 A design is a pair $(\mathcal{X}, \mathfrak{B})$ such that the following properties are
satisfied:
\begin{itemize}
\item $\mathcal{X}$ is a set of elements called points, and
\item $\mathfrak{B}$ is a collection of nonempty subsets of $\mathcal{X}$ called blocks.
\end{itemize}	\hfill $\square$
\end{definition}
A design is called $r$-regular if each point occurs in exactly $r$ blocks. A $r$-regular design containing $N$ points and $K$ blocks each of which has size $M$ is denoted by $r$- regular $(N,M,K)$ design. Clearly, we have $KM=rN$. In addition, a design is called $\eta$-cross if the intersection of any two different blocks has exactly $\eta$ points.  In this paper, we will use the following combinatorial structure to construct our schemes for Theorem \ref{th-design-CDC} and Theorem \ref{th-design-CDC-rs}.
\begin{definition}\rm(\cite{CD}, $t$-design)
\label{def-design}
Let $N$, $K$, $M$, $r$ and $t\geq 2$ and $\lambda$ be six  positive integers where $M<N$. A $t$-$(N,M,K,r,\lambda)$ design is a design $(\mathcal{X}, \mathfrak{B})$ where $\mathcal{X}$ has $N$ points and $\mathfrak{B}$ has $K$ blocks that satisfy
\begin{itemize}
\item $|\mathcal{B}|=M$ for any $\mathcal{B}\in \mathfrak{B}$;
\item every $t$-subset of $\mathcal{X}$ is contained in exactly $\lambda$ blocks;
\item every point occurs exactly in $r$ blocks, i.e., $r$-regular.
\end{itemize}	\hfill $\square$
\end{definition}

% {\blue The $t$-design will be used to construct our schemes for Theorem \ref{th-design-CDC} and Theorem \ref{th-design-CDC-rs}.}

By Definition \ref{def-design}, the number of blocks and the occurrence number of each point in blocks of $\mathfrak{B}$  are
\begin{align}\label{eq-value-b1}
K=\frac{\lambda{N\choose t}}{{M\choose t}},\ \ \ \ r= \frac{KM}{N}=\frac{\lambda \binom{N-1}{t-1}}{\binom{M-1}{t-1}},
\end{align} respectively. Here the value of the parameters $K$ and $r$ are determined by the parameters $t$, $N$, $M$ and $\lambda$. Obviously, any $t$-design is always regular. So a $t$-$(N,M,K,r,\lambda)$ design is also shorten as $t$-$(N,M,\lambda)$ design in this paper.   From \cite{CD},  we know that a $t$-$(N,M,\lambda)$-design is also a $t'$-$(N,M,\lambda_{t'})$ where $t'\leq t$ and
\begin{align}\label{eq--design-occurrence}
\lambda_{t'}= \frac{\lambda{N-t'\choose t-t'}}{ \binom{M-t'}{t-t'}},
\end{align} implying that a set of any $t'$ different points   occur exactly in $\lambda_{t'}$ blocks.

The $2$-$(N,M,\lambda)$ design is always called $(N,M,\lambda)$ balanced incomplete block design (in short BIBD).
%From \eqref{eq-value-b1} we can obtain that the number of blocks and the occurrence number of each point are $
%K=\frac{\lambda N(N-1)}{M(M-1)}$ and $r=\frac{KM}{N}=\frac{\lambda N(N-1)}{M(M-1)}\cdot M\cdot \frac{1}{N}=\frac{\lambda (N-1)}{M-1}$ respectively.
\begin{definition}\rm
\label{def-SBIBD}A BIBD in which $K=N$ (or, equivalently, $r=M$ or $\lambda(N-1)=
M(M-1)$) is called a symmetric BIBD (SBIBD).
\end{definition}It is worth noting that any SBIBD is regular and cross by the following well-known result.
\begin{lemma}[\cite{Stinson}]\rm
Suppose that $(\mathcal{X}, \mathfrak{B})$ is a $(N,M,\lambda)$ SBIBD. Then $|\mathcal{B}\cap \mathcal{B}'|=\lambda$ holds for all distinct blocks $\mathcal{B}$, $\mathcal{B}'\in \mathfrak{B}$.
		\hfill $\square$
\end{lemma}
\begin{example}\rm
\label{exam-SBIBD}
When $N=7$ and $M=3$, consider the design $(\mathcal{X},\mathfrak{B})$ where $\mathcal{X}=\{1,2,3,4,5,6,7\}$ and
\begin{align*}%\label{eq-blocks}
\mathfrak{B}=\{&\mathcal{B}_1=\{1,2,4\},\ \ \mathcal{B}_2=\{2,3,5\},\ \ \mathcal{B}_3=\{3,4,6\},\\
&\mathcal{B}_4=\{4,5,7\},\ \ \mathcal{B}_5=\{5,6,1\},\ \ \mathcal{B}_6=\{6,7,2\},\\
&\mathcal{B}_7=\{7,1,3\}\}.
\end{align*}
We can easily to verify that the design $(\mathcal{X},\mathfrak{B})$ is a $(7,3,1)$ SBIBD. So the cardinality of any two blocks in $\mathfrak{B}$ of the design  $(\mathcal{X},\mathfrak{B})$ is $1$.
\hfill $\square$
\end{example}

 Recently, P. Keevash in \cite{Peter} and  S. Glock et al., in \cite{GKLO} respectively proved the following existence conjecture for $t$-design.
\begin{lemma}[The existence conjecture for $t$-design\cite{Peter,GKLO}]\rm
\label{conjecture-design}
Given $t$, $M$ and $\lambda$, there exists an integer $N_0(t,M,\lambda)$, which is a function of $(t,M,\lambda)$,   such that
for any $N>N_0(t,M,\lambda)$, a $t$-$(N,M,\lambda)$ design exists if and only if for any $0\leq i\leq t-1$, the following condition holds
\begin{align}\label{ModCondition}
\lambda {N-i\choose t-i}\equiv\ 0\ \left({\text{mod}}\ {M-i\choose t-i}\right).
%\lambda {N-i\choose t-i}\ {\text{mod}}\ {M-i\choose t-i} \equiv\ 0.%\ \left(\right).
\end{align}		\hfill $\square$
\end{lemma}

The parameters in a $t$-$(N,M,\lambda)$ design should satisfy the condition \eqref{ModCondition} because of the third condition in Definition \ref{def-design}, i.e., every $t$-subset of $\mathcal{X}$ is contained in exactly $\lambda$ blocks. In fact, there exists another classic combinatorial structure called group divisible design (GDD), which relaxes the condition \eqref{ModCondition} and can also be used to construct our scheme in Theorem \ref{th-GDD-CDC}.
\begin{definition}\rm(\cite{CD}, $t$-GDD)
\label{def-GDD}
Let $M$, $t$, $q$ and $m$ be positive integers with $2\leq t\leq M\leq m$. A $(m ,q, M,\lambda)$ {\em group divisible $t$-design} ($t$-$(m ,q, M,\lambda)$ GDD) is a triple $(\mathcal{X}, \mathfrak{G}, \mathfrak{B})$ where \begin{itemize}
\item $\mathcal{X}$ is a set of $mq$ points,
\item $\mathfrak{G}=\{\mathcal{G}_1,\mathcal{G}_2,\ldots,\mathcal{G}_{m}\}$ is a partition of $\mathcal{X}$ into $m$ subsets each of which has size $q$ (called groups),
\item $\mathfrak{B}$ is a family of $M$-blocks of $\mathcal{X}$ such that every block intersects every group at most one point, and every $t$-subset of points from $t$ distinct groups belongs to exactly $\lambda$ blocks.
\end{itemize}		\hfill $\square$
\end{definition}
We can obtain the number of blocks and the occurrence number of each point in $\mathfrak{B}$ by Definition \ref{def-GDD} as
\begin{align}\label{eq-GDD-blocks-number}
K=\frac{\lambda{m\choose t}q^t}{{M\choose t}},\ \ \ \ r=\frac{KM}{N}=\frac{KM}{mq}=\frac{\lambda \binom{m-1}{t-1}q^{t-1}}{\binom{M-1}{t-1}},
\end{align}
respectively. It is well known that a $t$-$(m ,q, M,\lambda)$ GDD is also a $t'$-$(m ,q, M,\lambda_{t'})$ GDD where $t'\leq t$ and $\lambda_{t'}=\lambda \binom{m-t'}{t-t'}q^{t-t'}/\binom{k-t'}{t-t'}$ in \cite{CD}. In particular when $t'=2$ we have
\begin{align}\label{eq-GDD-occurrence}
\lambda_{2}= \frac{\lambda \binom{m-2}{t-2}q^{t-2}}{\binom{k-2}{t-2}},
\end{align}which implies that any two different points from different groups simultaneously occur in exactly $\lambda_2$ blocks.

\begin{example}\rm
\label{exam-1-t-GDD} When $m=M=3$ and $q=2$, let $\mathcal{X}=\{1,2,3,4,5,6\}$, groups
$$\mathfrak{G}=\{\mathcal{G}_1=\{1,2\}, \mathcal{G}_2=\{3,4\},  \mathcal{G}_3=\{5,6\}\}$$ and blocks
$\mathfrak{B}=\{ \mathcal{B}_1=\{1,3,5\}$, $ \mathcal{B}_2=\{1,4,6\}$, $\mathcal{B}_3=\{2,4,5\}$, $\mathcal{B}_4=\{2,3,6\}
\}$. Clearly $\mathfrak{G}$ is a partition of $\mathcal{X}$ and each group has $q=2$ points, i.e., the second condition of Definition \ref{def-GDD} holds. We can check that each pair of points that are from different groups occurs exactly in one block, i.e., the third condition of Definition \ref{def-GDD} holds. For instance, the pair $\{1,3\}$ occurs in block $\mathcal{B}_1$. So the above design $(\mathcal{X}, \mathfrak{G}, \mathfrak{B})$ is a $2$-$(3,2,3,1)$ GDD.
\hfill $\square$
\end{example}

By the above introductions of $t$-design and $t$-GDD, we can obtain the following observations.
\begin{remark}\rm
\label{remark-relation}
\begin{itemize}
\item The third property of $t$-GDD is similar to the second property of $t$-design, i.e., every $t$-subset of points (for a special condition in $t$-GDD) occurs exactly in $\lambda$ blocks.
\item  When $q=1$, any $t$-GDD is a $t$-design. When $q\geq 2$, from \eqref{eq-value-b1} and \eqref{eq-GDD-blocks-number}, we can not obtain the same parameters $K$ and $r$ for the same point set, block size $M$, and the parameters $t\geq 2$ and $\lambda$.
\end{itemize}
\end{remark}
There are many constructions and existences of the $t$-GDDs. For the detailed results, please see \cite[Section IV-4]{CD}.   Due to the fact that not every $t$-subset is contained by $\lambda$ blocks, the existence of the $t$-GDD is derived only for the parameters $t=2$ and $\lambda=1$. That is the following well-known result.
%The well known results are proposed by Moh{\'a}csy in \cite{Mohacsy-1,Mohacsy-2} in the following.
\begin{lemma}[The existence conjecture for $2$-GDD\cite{Chang}]\rm
 \label{lemma-existence-GDD}
Given $M$ and $m$ where $M\leq m$, there exists an integer $m_0(M,q)$, which is a function of $(M,q)$, such that
for any $m\geq m_0$, a $2$-$(m,q,M,1)$ GDD exists if and only if for any $0\leq i\leq 1$,
\begin{align}\label{eq-GDD-exsitence}
q^{t-i}{m-i\choose 2-i}\equiv\ 0\ \left(\text{mod}\ {M-i\choose 2-i}\right)
\end{align}
\end{lemma}

  %
%There are many constructions and existences of the $t$-designs and $t$-GDDs, especially on the results of BIBDs. For the detailed constructions,  please see \cite[Section II and IV]{CD}.

 Finally, we introduce the last useful concept which is a key transition turning combinatorial designs to our schemes.
 \begin{definition}[Dual design]\rm
\label{def-dual-design}
For any design $(\mathcal{X},\mathfrak{B})$,   the design $(\mathcal{V},\mathfrak{R})$ is called the dual design of  $(\mathcal{X},\mathfrak{B})$ if
\begin{itemize}
\item $\mathcal{V} =\mathfrak{B}$ and
\item $\mathfrak{R}=\{\mathcal{R}_x\ |\ x\in \mathcal{X}\}$ where $\mathcal{R}_x=\{\mathcal{B}\ |\ x\in \mathcal{B}, \mathcal{B}\in \mathfrak{B}\}$ for each $x\in \mathcal{X}$.\hfill $\square$
\end{itemize}

\end{definition}

By Definition \ref{def-dual-design},   the dual design of $(\mathcal{X},\mathfrak{B})$ is obtained by regarding the blocks $\mathfrak{B}$ as points and $\mathcal{X}$ as block set, where $\mathcal{B}\in \mathfrak{B}$ is contained by $x\in \mathcal{X} $ if and only if $x\in \mathcal{B}$.  Clearly, a design is a dual design of its dual design.

 For any $t$-design and $t$-GDD, their dual designs have the following useful properties, whose proof is given in Appendix \ref{appendix-dual-design}.

%It is not difficult to obtain the following {results whose proof are included in Appendix \ref{appendix-dual-design} and Appendix \ref{appendix-dual-GDD}.}
\begin{lemma}[Regular and cross design via $t$-design]\rm
\label{lemm-dual-design-system}
The dual design of a $t$-$(N,M,\lambda)$ design is a design where
\begin{enumerate}
\item there are exactly $K=\lambda{N\choose t}/{M\choose t}$ points each of which occurs in exactly $M$ blocks;
\item there are $N$ blocks each of which has size $r=\lambda{N-1\choose t-1}/ \binom{M-1}{t-1}$;
\item any $t'$ distinct blocks intersect in exactly $\lambda_{t'}=\lambda{N-t'\choose t-t'}/ \binom{M-t'}{t-t'}$ points where $2\leq t'\leq t$. 	\hfill $\square$
\end{enumerate}	
\end{lemma}

\begin{lemma}[Regular and almost cross design via $t$-GDD]\rm
\label{lemm-dual-GDD-system}
The dual design of a $t$-$(m ,q, M,\lambda)$ GDD is a design where
\begin{enumerate}
\item there are exactly $K=\lambda{m\choose t}q^t/{M\choose t}$ points each of which occurs in exactly $M$ blocks;
\item there are $mq$ blocks each of which has size $r=\lambda{m-1\choose t-1}q^{t-1}/{M-1\choose t-1}$;
\item there are $m{q\choose 2}$ pairs of distinct blocks whose intersection is an empty set, i.e., the two different points in the same group of the GDD;
\item there are ${mq\choose 2}-m{q\choose 2}$ pairs of distinct blocks whose intersection has exactly $\lambda_2=\lambda{m-2\choose t-2}q^{t-2}/\binom{M-2}{t-2}$ points, i.e., the number of pairs containing the points included in the block of the GDD.	\hfill $\square$
\end{enumerate}
\end{lemma}% low-complexity and communiation-efficient

Let us take the $(7,3,1)$ SBIBD in Example \ref{exam-SBIBD} and the $2$-$(3,2,3,1)$ GDD in Example \ref{exam-1-t-GDD} to further introduce the concept of dual design and their properties in Lemma \ref{lemm-dual-design-system} and Lemma \ref{lemm-dual-GDD-system}, respectively.

\begin{example}\rm
\label{exam-dual-design}
By Definition \ref{def-dual-design} we can obtain the dual design $(\mathcal{V},\mathfrak{R})$ of the $(7,3,1)$ SBIBD in Example \ref{exam-SBIBD} where point set $
\mathcal{V}=\mathfrak{B}=\{\mathcal{B}_1,\mathcal{B}_2,
\mathcal{B}_3,\mathcal{B}_4,\mathcal{B}_5,\mathcal{B}_6,
\mathcal{B}_7\}$ and block set
\begin{align}
\mathfrak{R}=\{&
\mathcal{R}_1=\{\mathcal{B}_1, \mathcal{B}_5,\mathcal{B}_7\},
\mathcal{R}_2=\{\mathcal{B}_1,\mathcal{B}_2,\mathcal{B}_6\},\nonumber\\
&\mathcal{R}_3=\{\mathcal{B}_2,\mathcal{B}_3,\mathcal{B}_7\},
\mathcal{R}_4=\{\mathcal{B}_1,\mathcal{B}_3,\mathcal{B}_4\},\nonumber\\
&\mathcal{R}_5=\{\mathcal{B}_2,\mathcal{B}_4,\mathcal{B}_5\},
\mathcal{R}_6=\{\mathcal{B}_3,\mathcal{B}_5,\mathcal{B}_6\},\nonumber\\
&\mathcal{R}_7=\{\mathcal{B}_4,\mathcal{B}_6,\mathcal{B}_7\}\}.\label{eq-dual-blocks}
\end{align}
%{\blue We can see that each block of the dual design has exactly $r=3$ points which the occurrence number of each point in blocks of $\mathfrak{B}$; each point occurs exactly $M=3$ blocks which is exactly each block size of $\mathfrak{B}$; the intersection of any two distinct blocks has exactly $\lambda=1$ points which corresponds to the occurrence number of any distinct points in blocks of $\mathfrak{B}$.
%
%In fact for any $t$-design, the above claim also holds, i.e., the following result.}
We can check that $|\mathcal{V}|=K=\lambda{N\choose t}/{M\choose t}={7\choose 2}/{3\choose 2}=7$ and each block has $r=\lambda{N-1\choose t-1}/\binom{M-1}{t-1}={7-1\choose 2-1}/\binom{3-1}{2-1}=3$ points, i.e., the first two properties of Lemma \ref{lemm-dual-design-system} hold; the intersection of any two different blocks of $\mathfrak{R}$ has exactly $\lambda=1$ point, i.e., the third property of Lemma \ref{lemm-dual-design-system} holds.  In addition, we can easily verify that the designs $(\mathcal{V},\mathfrak{R})$ is also a $(7,3,1)$ SBIBD.
\hfill $\square$
\end{example}
\begin{example}\rm
\label{exam-t-dual-GDD}
 By Definition \ref{def-dual-design} we can obtain the dual design $(\mathcal{V},\mathfrak{R})$ of $2$-$(3,2,3,1)$ GDD in Example \ref{exam-1-t-GDD} where the point set $
\mathcal{V}=\{\mathcal{B}_1,\mathcal{B}_2,\mathcal{B}_3,\mathcal{B}_4\}$ and the block set
\begin{align}
\mathfrak{R}=\{
&\mathcal{R}_1=\{\mathcal{B}_1,\mathcal{B}_2\},\
\mathcal{R}_2=\{\mathcal{B}_3,\mathcal{B}_4\},\nonumber\\
&\mathcal{R}_3=\{\mathcal{B}_1,\mathcal{B}_4\},\
\mathcal{R}_4=\{\mathcal{B}_2,\mathcal{B}_3\},\nonumber \\ &\mathcal{R}_5=\{\mathcal{B}_1,\mathcal{B}_3\},\
    \mathcal{R}_6=\{\mathcal{B}_2,\mathcal{B}_4\}\}.\label{eq-dual-GDD-blocks}
\end{align}
   We can check that $|\mathcal{V}|=K=\lambda{m\choose t}q^t/{M\choose t}={3\choose 2}2^2/{3\choose 2}=4$ and each block has $r=\lambda{m-1\choose t-1}q^{t-1}/{M-1\choose t-1}={3-1\choose 2-1}2^{2-1}/{3-1\choose 2-1}=2$ points, i.e., the first two properties of Lemma \ref{lemm-dual-GDD-system} hold; the intersection of any two blocks labeled by groups $\mathcal{G}_1=\{1, 2\}$, $\mathcal{G}_2=\{3,4\}$ and $\mathcal{G}_3=\{5,6\}$ in Example \ref{exam-1-t-GDD} is an empty set, i.e., the third property of Lemma \ref{lemm-dual-GDD-system} holds; and the intersection of any two blocks with labels choosing from two of groups $\mathcal{G}_1$, $\mathcal{G}_2$ and $\mathcal{G}_3$ contains exactly $\lambda=1$ point, i.e., the fourth property of Lemma \ref{lemm-dual-GDD-system} holds. For instance, the intersection of blocks $\mathcal{R}_1$ and $\mathcal{R}_2$ is an empty set, and the intersection of blocks $\mathcal{R}_1$ and $\mathcal{R}_4$ is $\{\mathcal{B}_2\}$.
\hfill $\square$
\end{example}

As we will see in Section \ref{Sec:Schemes},  we will use the dual design to construct coding schemes based on Lemma \ref{lemm-dual-design-system} and Lemma \ref{lemm-dual-GDD-system}, obtaining schemes for the case $r=s$ in Theorem \ref{th-design-CDC} and Theorem \ref{th-GDD-CDC}, respectively. Moreover, the two Lemmas are also helpful to construct schemes for the case $r\neq s$. For instance, by taking $t'=t-1$ in the third statement of Lemma  \ref{lemm-dual-design-system}, we propose our last scheme in Section \ref{sec-extension} for Theorem \ref{th-design-CDC-rs}. In fact, we can obtain other schemes with $r\neq s$ by choosing any other value of $t'\in [2,t-2]$. In addition, a similar approach can be applied to obtain a $t$-GDD scheme with  $r\neq s$, and is omitted due to page limits. %Since the constructing method proposed in Theorem \ref{th-GDD-CDC}, in this paper we only consider the intersection of any two different blocks in  Lemma \ref{lemm-dual-GDD-system}.
   %Specifically, we will use the first two statements of Lemma \ref{lemm-dual-design-system} and Lemma \ref{lemm-dual-GDD-system} to derive the values of the parameters $K$, $r$ and $s$ of our new schemes, and use the last statements, i.e., the last statement of Lemma \ref{lemm-dual-design-system} and the last two statements of Lemma \ref{lemm-dual-GDD-system}, to design the delivery strategies of our new schemes. Finally, in order to show that our constructing method also works for the case $r\neq s$, we propose our last scheme in Theorem \ref{th-design-CDC-rs} by taking $t'=t-1$ in the third statement of Lemma  \ref{lemm-dual-design-system} as an example.

%By the first and third property in Lemma \ref{lemm-dual-design-system}, the dual design of any $t$-$(N,M,\lambda)$ design is $M$-regular and $\lambda_2$-cross.
%Similar to Lemma \ref{lemm-dual-design-system}, the following result can be obtained. The proof is included in Appendix .

\section{Main Results}
\label{sec-main}
In this section, we will first present the results of our two schemes based on  $t$-design and $t$-GDD, respectively for the case  $r=s$, and then show that our new schemes are asymptotically optimal under one-shot linear delivery when $r+s\leq K$. Finally, we compare our schemes with that of the state-of-art schemes, and show that ours have  smaller file number $N$, output function number $Q$, and communication load.

Given a $t$-design $(\mathcal{X}, \mathfrak{B})$,  we can obtain its dual design $(\mathcal{V},\mathfrak{R})$. By taking the point set   $\mathcal{V}$ as the worker set and the block set $\mathfrak{R}$ to generate the data placement and output function assignment, and using the cross property of $(\mathcal{V},\mathfrak{R})$ to generate the delivery strategy for the intermediate values, we obtain the following Theorem, whose detailed proof is given in  Section \ref{sub-proof-th-Design-CDC}.

\begin{theorem}[$t$-design scheme]\rm
\label{th-design-CDC}
 If there exists a $t$-$(N,M,\lambda)$ design, we can obtain a $(K=\lambda{N\choose t}/{M\choose t}$, $r=\lambda{N-1\choose t-1}/ \binom{M-1}{t-1}$, $s=r$, $N$, $Q=N)$ cascaded CDC scheme with the communication load $L_{t\text{-Design}}={(N-1)}/{(2N)}$.
		\hfill $\square$
\end{theorem}

From Lemma \ref{conjecture-design} and Theorem \ref{th-design-CDC}, we can obtain an arbitrary  $(K=\lambda{N\choose t}/{M\choose t}$, $r=\lambda{N-1\choose t-1}/ \binom{M-1}{t-1}$, $s=r$, $N$, $Q=N)$ cascaded CDC scheme with the communication load $L_{t\text{-Design}}=\frac{N-1}{2N}$ for any parameters $t$, $M$ and $\lambda$ when $N$ is larger than or equal to $N_0(t,M,\lambda)$ and satisfies \eqref{ModCondition}. 
To relax the parameter restrictions in \eqref{ModCondition}, we propose a new scheme based $t$-GDD  that is feasible for some parameters $K$ and $N$ that $t$-design scheme can not achieve (See Table \ref{tab-Main-results}). The result is presented in the following Theorem listed in Table \ref{tab-Main-results}, whose proof is given in  Section \ref{sub-proof-th-GDD-CDC}.

%Then by Theorem \ref{th-design-CDC}, we can obtain the $(K=\lambda{N\choose t}/{M\choose t}$, $r=\lambda{N-1\choose t-1}/ \binom{M-1}{t-1}$, $s=r$, $N$, $Q=N)$ cascaded CDC scheme with the communication load $L_{t\text{-Design}}=\frac{N-1}{2N}$ for any parameters $t$, $M$ and $\lambda$ when $N\geq N_0(t,M,\lambda)$.

%{\blue Similar to the construction of the scheme in Theorem \ref{th-design-CDC}, in the following we can obtain the result of our second scheme based $t$-GDD. The detailed proof is given in \ref{sub-proof-th-GDD-CDC}.}

%{\red
%{\bf Could we delete the following paragraph?}
%
%
%According to the proof of Theorem \ref{th-design-CDC} in Subsection \ref{sub-proof-th-Design-CDC}, we can see that the key point of designing the delivery strategy is the second property of {\blue dual design $(\mathcal{V},\mathfrak{R})$ which is derived by the second property of the $t$-design $(\mathcal{X},\mathfrak{B})$.} This implies that when a design has a similar property to the second property of $t$-design, we can also obtain a new scheme based on such design. For instance, given a $t$-GDD $(\mathcal{X}, \mathfrak{G}, \mathfrak{B})$, by taking the block set $\mathfrak{B}$ as worker set and the point set $\mathcal{X}$ to generate the data placement and output function assignment, and using the third property of $t$-GDD to generate the delivery strategy for the intermediate values, the following Theorem can be obtained, whose proof is given in \ref{sub-proof-th-GDD-CDC}.}

\begin{theorem}[$t$-GDD scheme]\rm
\label{th-GDD-CDC}
 If there exists a $t$-$(m ,q, M,\lambda)$ GDD, we can obtain a $(K=\lambda{m\choose t}q^t/{M\choose t}$, $r=\lambda q^{t-1}\binom{m-1}{t-1}/\binom{M-1}{t-1}$, $ s=r$, $N=mq$, $Q=mq)$ cascaded CDC scheme with the communication load $L_{t\text{-GDD}}= \frac{1}{2}+\frac{q-2}{2mq}$.
		\hfill $\square$
\end{theorem}
%By Lemma \ref{lemma-existence-GDD} and Theorem  \ref{th-design-CDC}, we can obtain a $(K={m\choose t}q^t/{M\choose t}$, $r= q(m-1)/(m-1)$, $ s=r$, $N=mq$, $Q=mq)$ cascaded CDC scheme with the communication load $L_{2\text{-GDD}}= \frac{1}{2}+\frac{q-2}{2mq}$ for any parameters $m$, $q$ and $M$ where $m$ is larger than or equal to $m_0(M,q)$ and satisfies \eqref{eq-GDD-exsitence}, i.e., the scheme in Table \ref{tab-Main-results}.
From Table \ref{tab-Main-results}, we can see that by letting $m=N$, $q=1$, the $2$-GDD scheme shares the same system parameters $(K,r,s,N,Q)$ as that of Theorem  \ref{th-design-CDC}. But when $q>1$, the scheme in Theorem \ref{th-GDD-CDC} can achieve some other parameters $K$, $r=s$ for the same $N$ and $Q$. In the following subsection, we compare schemes of  Theorem  \ref{th-design-CDC} and  Theorem \ref{th-GDD-CDC} with the state-of-art works in \cite{{LMYA},JWZ,WCJ,JQ}.

%In Section \ref{Sec:Schemes}, we can see that $t$-GDD scheme can be regarded as the generalization of the $t$-design.

\subsection{The  Asymptotical Optimality under One-shot Linear Delivery}
\label{sub:asymptotical}
For any scheme under one-shot linear delivery, we can derive the following maximum multicast gain whose proof is included in Appendix \ref{proofOptimal}.
\begin{lemma}\rm%[The maximum multicast gain under {\blue one-shot   linear} coding  delivery]
\label{lem-maximum-gain}
For the $(K,r,s,N,Q)$ cascaded CDC scheme under one-shot linear delivery, the maximum multicast gain
% is upper bounded by
$g^*_\text{one-shot}\leq  \min\{r+s-1,K-1\}$. 		\hfill $\square$
\end{lemma}

By  \eqref{eq-communication-load} and Lemma \ref{lem-maximum-gain}, for any $(K,r,s,N,Q)$ cascaded CDC scheme under one-shot linear delivery and each worker stores $M$ files, the minimum communication load, denoted by $L_{\text{one-shot}}^{*}(r,s)$, is lower bounded by
\begin{align}\label{eq-lower-boud}
L_{\text{one-shot}}^{*}(r,s)\geq \frac{s(1-M/N)}{ \min\{r+s-1,K-1\}}.
\end{align}

 In the following, we only focus on the case $r+s\leq K$. By Remark \ref{remark-coded-gain} in Subsection \ref{sub-proof-th-Design-CDC}, we have that the $t$-design scheme achieves a multicast gain $g=2(r-\lambda_2)$, which is $2\lambda_2-1$ less than the upper bound of one-shot maximum multicast gain $2r-1$ when $r=s$. Especially, when we use the BIBD with $\lambda=1$, we have the following result.
\begin{proposition}\rm
\label{pro-max-gian}
The scheme generated by a $(N,M,\lambda=1)$ BIBD has a multicast gain $g=2r-2$, which is just one less than the upper bound of the maximum one-shot multicast gain.		\hfill $\square$
\end{proposition}

% As we will see in Remark \ref{remark-coded-gain} in Subsection \ref{sub-proof-th-Design-CDC}, the $t$-design scheme generated by a $(N,M,1)$ BIBD can achieve the multicast gain $g=2r-2$ that is just one less than the upper bound of  $g^*_\text{one-shot}$.

 Moreover,
from   \eqref{eq-communication-load}, Lemma \ref{lem-maximum-gain}, Theorem  \ref{th-design-CDC} and Theorem \ref{th-GDD-CDC}, we can prove the asymptotical optimality of our schemes. We formally present this result in the following Corollary, whose proof is given in Appendix \ref{AppendixAsyOpt}.
\begin{corollary}[ Asymptotical optimality under one-shot linear delivery]\rm\label{orderOptimal} When $r+s<K$, we have that
\begin{itemize}
\item the $t$-design scheme in Theorem \ref{th-design-CDC} achieves  one-shot asymptotically optimal communication load if $K\gg r$ and $K\rightarrow\infty,$   for any give values of $M$, $t$ and $\lambda$;
\item the $t$-GDD scheme in Theorem \ref{th-GDD-CDC} achieves one-shot asymptotically optimal communication load if $m$ is large for any given values of $q$, $M$, $t$ and $\lambda$. 		\hfill $\square$
\end{itemize}
\end{corollary}

%
%Especially, when we use the BIBD with $\lambda=1$, the obtained scheme in Theorem \ref{th-design-CDC} has the multicast gain $2(r-1)$, which is just one less than the maximum one-shot multicast gain. The following Corrollary shows  that the scheme in Theorem \ref{th-design-CDC} is asymptotically optimal under one-shot delivery, whose proof is given in Appendix \ref{AppendixAsyOpt}.
%
%
%
%
%
%
%So the minimum communication load for a  $(K,r,s,N,Q)$ cascaded CDC scheme under one-shot delivery is lower bounded by $L_{\text{one-shot}}^{*}(r,s)\geq \frac{s(1-M/N)}{r+s-1}$.  By Lemma \ref{lem-maximum-gain} and Remark \ref{remark-coded-gain} in Subsection \ref{sub-proof-th-Design-CDC}, we have that the $t$-design scheme achieves a multicast gain $g=2(r-\lambda_2)$, which is $2\lambda_2-1$ less than the upper bund of one-shot maximum multicast gain $2r-1$ when $r=s$. Especially, when we use the BIBD with $\lambda=1$, the obtained scheme in Theorem \ref{th-design-CDC} has the multicast gain $2(r-1)$, which is just one less than the maximum one-shot multicast gain. The following Corrollary shows  that the scheme in Theorem \ref{th-design-CDC} is asymptotically optimal under one-shot delivery, whose proof is given in Appendix \ref{AppendixAsyOpt}.
\subsection{Comparisons with The State-of-art Works}
\label{subsection-comparisions-state}
\begin{table*}[tb]
  \centering
  \renewcommand\arraystretch{1}
    \setlength{\tabcolsep}{1mm}{
  \caption{New schemes via $(N,M,1)$ BIBD and $2$-$(m,q,M,1)$ GDD satisfying  $2<M<N$, $M<m$, ${M-i\choose 2-i}|{N-i\choose 2-i}$, ${M-i\choose 2-i}|q^{2-i}{m-i\choose 2-i}$ where $i=0,1$.  \label{tab-new-design-t=2} }
  \begin{tabular}{|c|c|c|c|c|c|c|c|}
\hline
Schemes &\tabincell{c}{Worker number\\  $K$} & \tabincell{c}{Computation \\  Load $r$}& \tabincell{c}{Replication \\ Factor $s$}&  \tabincell{c}{Number of \\ Files $N$}   & \tabincell{c}{Number of Reduce \\ Functions $Q$}&  \tabincell{c}{Communication \\ Load $L_{\text{comm.}}$}& \tabincell{c}{Operation \\ Field $\mathbb{F}_2$}\\
\hline
In Theorem \ref{th-design-CDC} &
$\frac{ N(N-1)}{M(M-1)}$ & $\frac{N-1}{M-1}$ & $\frac{N-1}{M-1}$ & $N$ & $N$& $\frac{N-1}{2N}<\frac{1}{2}$&Yes\\ [4pt] \hline
In Theorem \ref{th-GDD-CDC} &$\frac{ m(m-1)q^2}{M(M-1)}$& $\frac{(m-1)q}{M-1}$&$\frac{ (m-1)q}{M-1}$& $mq$ &$mq$&$\frac{1}{2}+\frac{q-2}{2mq}$& Yes\\
\hline
\end{tabular}
}
\end{table*}
\begin{table*}[tb]
  \centering
  \renewcommand\arraystretch{1}
    \setlength{\tabcolsep}{1mm}{
  \caption{Existing cascaded CDC schemes where the positive integers $M_1$, $N_1$ and $\lambda$ satisfying  $2<M_1<N_1$, ${M_1-i\choose 2-i}|{N_1-i\choose 2-i}$ and $M_1|N_1$ where $i=0,1$. \label{tab-known-design-CDC-t=2} }
  \begin{tabular}{|c|c|c|c|c|c|c|c|}
\hline
Schemes &\tabincell{c}{Worker number\\  $K$} & \tabincell{c}{Computation \\  Load $r$}& \tabincell{c}{Replication \\ Factor $s$}&  \tabincell{c}{Number of \\ Files $N$}   & \tabincell{c}{Number of Reduce \\ Functions $Q$}&  \tabincell{c}{Communication \\ Load $L_{\text{comm.}}$}& \tabincell{c}{Operation \\ Field $\mathbb{F}_2$}\\
\hline
\cite{WCJ}
&\multirow{3}{*}{\tabincell{c}{\  \\ \ \\ $\frac{ N_1(N_1-1)}{M_1(M_1-1)}$}}
&\multirow{3}{*}{ \tabincell{c}{\  \\ \ \\ $\frac{N_1-1}{M_1-1}$}}
&\multirow{3}{*}{\tabincell{c}{\  \\ \ \\  $\frac{N_1-1}{M_1-1}$}}
&$(\frac{N_1}{M_1})^{\frac{N_1-1}{M_1-1}}$
&$(\frac{N_1}{M_1})^{\frac{N_1-1}{M_1-1}}$
&\tabincell{c}{$\frac{1}{2}-\frac{1}{2}\cdot\left( \left(\frac{M_1}{N_1}\right)^{\frac{N_1-1}{M_1-1}} -\right.$\\
$\left.\left(1\!-\!\frac{M_1}{N_1}\right)^{\frac{N_1-1}{M_1-1}} \!\cdot\! \frac{1}{\frac{2(N_1-1)}{M_1-1}-1}\right)$}&No\\
\cline{1-1} \cline{5-8}
\cite{JWZ} &
& &&$\frac{ N_1(N_1-1)}{M_1(M_1-1)}$& $\frac{ N_1(N_1-1)}{M_1(M_1-1)}$&\multirow{2}{*}{\tabincell{c}{\  \\
$\frac{M_1}{M_1-1}\cdot\frac{N_1-M_1}{N_1}$} }&No\\[0.3cm]
\cline{1-1} \cline{5-6}\cline{8-8}
\cite{JQ}&
& &&$(\frac{N_1}{M_1})^{\frac{N_1-1}{M_1-1}-1}$& $\frac{N_1}{M_1}$& &Yes\\[0.3cm]
\hline
\end{tabular}
}
\end{table*}
We first present the following Corollary that compares the schemes of Theorem  \ref{th-design-CDC} and Theorem \ref{th-GDD-CDC}  with the vanilla CDC scheme in \cite{LMYA},  whose proof is included in Appendix \ref{secConverse-cororllary-2}.

%Now we compare the communication loads in  Theorem  \ref{th-design-CDC} and Theorem \ref{th-GDD-CDC} with that of the state-of-art works in \cite{{LMYA},JWZ,WCJ,JQ}.
%By Table \ref{tab-known-CDC}, we first compare the schemes of Theorems  \ref{th-design-CDC} and \ref{th-GDD-CDC} with the CDC scheme in \cite{LMYA}. The comparison results are given in the following Corrolary, whose proof is included in Appendix \ref{secConverse}.
 \begin{corollary}\label{CorollaryThm1vsCDC}%[Asymptotic Optimality]
 \rm
\begin{itemize}
\item For the cascaded CDC system described in Section \ref{secSystem} with computation load $r=s$ and $N=Q$, the schemes for Theorem \ref{th-design-CDC} and Theorem \ref{th-GDD-CDC} have much smaller $N$ and $Q$ than that of \cite{LMYA}.
\item
 If $K\gg r$ and $K\rightarrow\infty,$ we  have communication loads $L_{t\text{-Design}}<1/2$,  $L_{t\text{-GDD}}\geq 1/2$, and $L_{\text{LMYA}}>1/2$, indicating $L_{t\text{-GDD}}>L_{\text{LMYA}}>L_{t\text{-Design}}$. Moreover,
\begin{itemize}
\item If $K\gg r, \ K\rightarrow\infty,$ and  $N\rightarrow \infty$, then  $L_{t\text{-Design}}=L_{\text{LMYA}}=\frac{1}{2}$.
\item If  $m\to \infty$ and $m\gg q $, then      $L_{t\text{-GDD}}=L_{\text{LMYA}}=\frac{1}{2}$.
\end{itemize}		\hfill $\square$
\end{itemize}
 \end{corollary}
 \begin{remark}\rm\label{RemarkImprove}
 In  \cite{WCJ}, the authors showed that their scheme achieves a strictly smaller communication load than \cite{LMYA} when $r=s=2$. However, the numerical results in \cite[Section V-A]{WCJ} showed that the improvement no longer holds when $r=s>2$.   Remarkably, Corollary \ref{CorollaryThm1vsCDC} shows that when $K$ is sufficiently large such that $K\gg r$, our schemes always improve over the scheme in \cite{LMYA} for arbitrary $r=s$.
 \hfill $\square$
 \end{remark}

Next, we compare our schemes with the state-of-art schemes \cite{JWZ,WCJ,JQ} that aim to relax the implementation limitation of CDC scheme \cite{LMYA} that requires exponentially large numbers of
both input files and output functions.

 Firstly, by setting $\lambda=1$ and $t=2$ we obtain the system parameters and communication loads of Theorem  \ref{th-design-CDC} and Theorem \ref{th-GDD-CDC} as shown in  Table \ref{tab-new-design-t=2}.

%By Definition \ref{def-SBIBD}, a SBIBD is a special $2$-design with the point number equal to the block number, i.e., $N=K$. This implies that we can obtain the scheme with more flexible parameters compared with the scheme in \cite{JWZ}. So let us first compare with the scheme in \cite{JWZ}.
%
%
% Note that by Definition \ref{def-design}, we know that SBIBD is a special $2$-design with the point number equal to the block number, i.e., $N=K$.
%
%
%
%
% \vskip 3cm
 To compare our schemes with  that of \cite{WCJ,JWZ,JQ}, we set $(K=\frac{N_1(N_1-1)}{M_1(M_1-1)}$,   $r=s=\frac{N_1-1}{M_1-1})$ and  $(K=\frac{ m(m-1)q^2}{M(M-1)}$, $r=s=\frac{(m-1)q}{M-1})$ in  Table \ref{tab-known-CDC}, and obtain Table \ref{tab-known-design-CDC-t=2} and Table \ref{tab-known-GDD-CDC-t=2} respectively.  From Table \ref{tab-new-design-t=2}, Table \ref{tab-known-design-CDC-t=2} and Table \ref{tab-known-GDD-CDC-t=2}, we obtain the following corollaries, whose proofs are given in Appendix \ref{secConverse-cororllary-3-4}.
\begin{corollary}\rm
\label{CorollaryThm1vsOthers}
For the same parameters $K$, $r=s$ our scheme in Theorem \ref{th-design-CDC} has
\begin{itemize}
\item much smaller file number $N$ and output function number $Q$, and smaller communication load than that of the scheme in \cite{WCJ} if $  {K}/{r}>1+(2r+1)^{ {1}/{r}}$ holds\footnote{In fact this condition   holds for most of positive integer $r$ since the limit of the function $f(x)=x^{\frac{1}{x}}$ is 1.};
\item smaller file number $N$ and output function number $Q$, and smaller communication load than that of the scheme in \cite{JWZ}. In addition, a SBIBD is a special $2$-design with the point number equal to the block number, i.e., $N=K$, by Definition \ref{def-SBIBD}. This implies that we can obtain the scheme with more flexible parameters compared with the scheme in \cite{JWZ};
\item smaller file number $N$, some larger output function number $Q$, and smaller communication load than that of the scheme in \cite{JQ}  if $N_1\geq 3M_1$ and $M_1\geq 9$.\hfill $\square$
\end{itemize}		

\end{corollary}

%Now let us consider comparisons between our scheme in Theorem \ref{th-GDD-CDC} and schemes in \cite{WCJ,JWZ,JQ} respectively when the parameters $K=\frac{ m(m-1)q^2}{M(M-1)}$ and $r=s=\frac{(m-1)q}{M-1}$. Then the schemes in \cite{WCJ,JWZ,JQ}  can be obtained in Table \ref{tab-known-GDD-CDC-t=2}.
\begin{table*}
  \centering
  \renewcommand\arraystretch{1}
    \setlength{\tabcolsep}{1mm}{
  \caption{Existing cascaded CDC schemes where the positive integers $M$, $N$ and $M<m$ satisfying $2<M<m$,
${M-i\choose 2-i}|q^{2-i}{m-i\choose 2-i}$ and $M|mq$ where $i=0,1$.
  \label{tab-known-GDD-CDC-t=2} }
  \begin{tabular}{|c|c|c|c|c|c|c|c|}
\hline
Schemes &\tabincell{c}{Worker number\\  $K$} & \tabincell{c}{Computation \\  Load $r$}& \tabincell{c}{Replication \\ Factor $s$}&  \tabincell{c}{Number of \\ Files $N$}   & \tabincell{c}{Number of Reduce \\ Functions $Q$}&  \tabincell{c}{Communication \\ Load $L_{\text{comm.}}$}& \tabincell{c}{Operation \\ Field $\mathbb{F}_2$}\\
\hline
\cite{WCJ} &\multirow{3}{*}{\tabincell{c}{\  \\ \ \\ $\frac{ m(m-1)q^2}{M(M-1)}$}}
&\multirow{3}{*}{ \tabincell{c}{\  \\ \ \\ $\frac{(m-1)q}{M-1}$}}
&\multirow{3}{*}{\tabincell{c}{\  \\ \ \\  $\frac{(m-1)q}{M-1}$}}
& $(\frac{mq}{M})^{\frac{(m-1)q}{M-1}-1}$
&$(\frac{mq}{M})^{\frac{(m-1)q}{M-1}-1}$&
\tabincell{c}{$\frac{1}{2}-\frac{1}{2}\cdot\left( \left(\frac{M}{mq}\right)^{\frac{(m-1)q}{M-1}} +\right.$\\
$\left.\left(1-\frac{M}{mq}\right)^{\frac{(m-1)q}{M-1}} \cdot \frac{1}{\frac{2(m-1)q}{M-1}-1}\right)$}&No\\
\cline{1-1} \cline{5-8}

\cite{JWZ} &
& &&$\frac{(m-1)q}{M-1}$& $\frac{(m-1)q}{M-1}$&\multirow{2}{*}{\tabincell{c}{\  \\
$\frac{\frac{(m-1)q}{M-1}}{\frac{(m-1)q}{M-1}-1}
\cdot\left(1-\frac{M}{mq}\right)>1-\frac{M}{mq}$} }&No\\[0.3cm]
\cline{1-1} \cline{5-6}\cline{8-8}
\cite{JQ}&
& &&$(\frac{mq}{M})^{\frac{(m-1)q}{M-1}-1}$& $\frac{mq}{M}$& &Yes\\[0.3cm]
\hline
\end{tabular}
}
\end{table*}

\begin{corollary}\rm
\label{remark-performance-GDD-t=2}
For the same parameters $K$, $r=s$ our scheme in Theorem \ref{th-GDD-CDC} has
\begin{itemize}
\item much smaller file number $N$ and output function number $Q$  than that of the scheme in \cite{WCJ} with the approximately same communication load;
\item has smaller file number $N$ and output function number, $Q$,  and smaller communication load that that of the scheme in \cite{JWZ};
\item smaller file number $N$ and some larger output function number $Q$ and smaller communication load that that of the scheme in \cite{JQ}.\hfill $\square$
\end{itemize}		

\end{corollary}

\begin{figure}[h]
	\centering
	\includegraphics[scale=0.5]{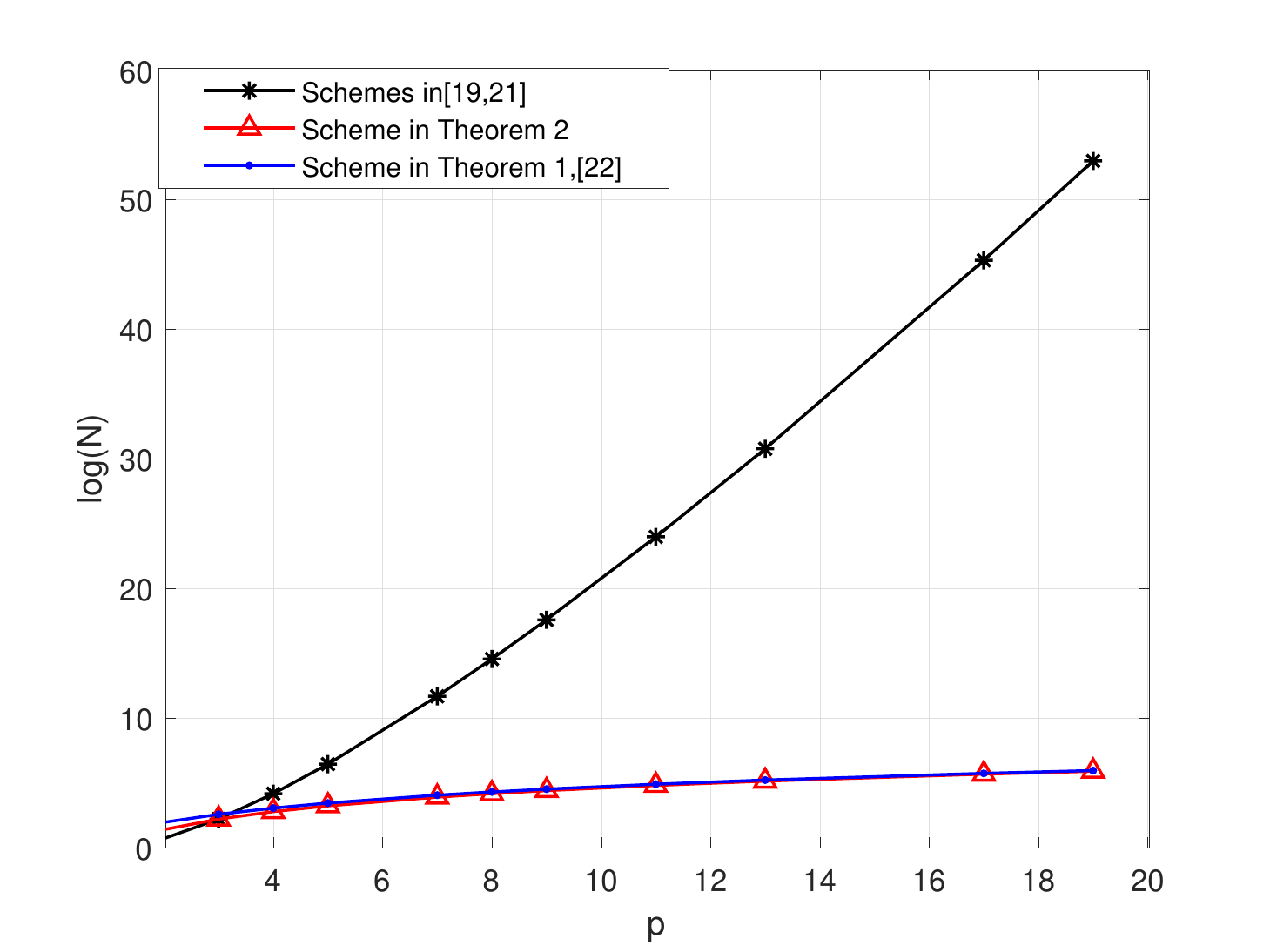}
	\caption{Numbers of files of the schemes in Theorems \ref{th-design-CDC} and \ref{th-GDD-CDC} and the existing schemes.}
	\label{fig:N}
\end{figure}

\begin{figure}[h]
	\centering
	\includegraphics[scale=0.5]{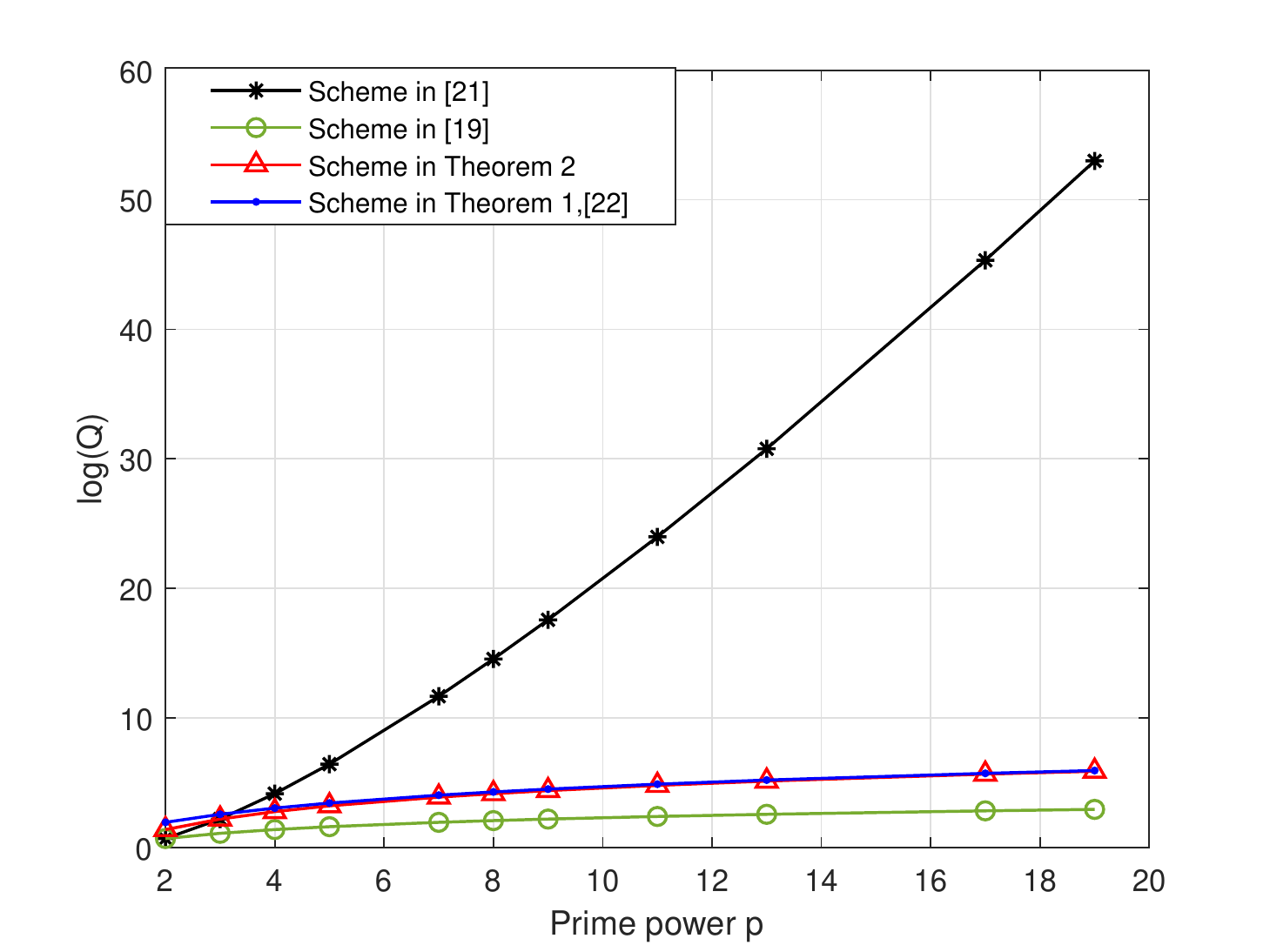}
	\caption{Numbers functions of the schemes in Theorems \ref{th-design-CDC} and \ref{th-GDD-CDC} and texisting schemes.}
	\label{fig:Q}
\end{figure}

\begin{figure}[h]
	\centering
	\includegraphics[scale=0.5]{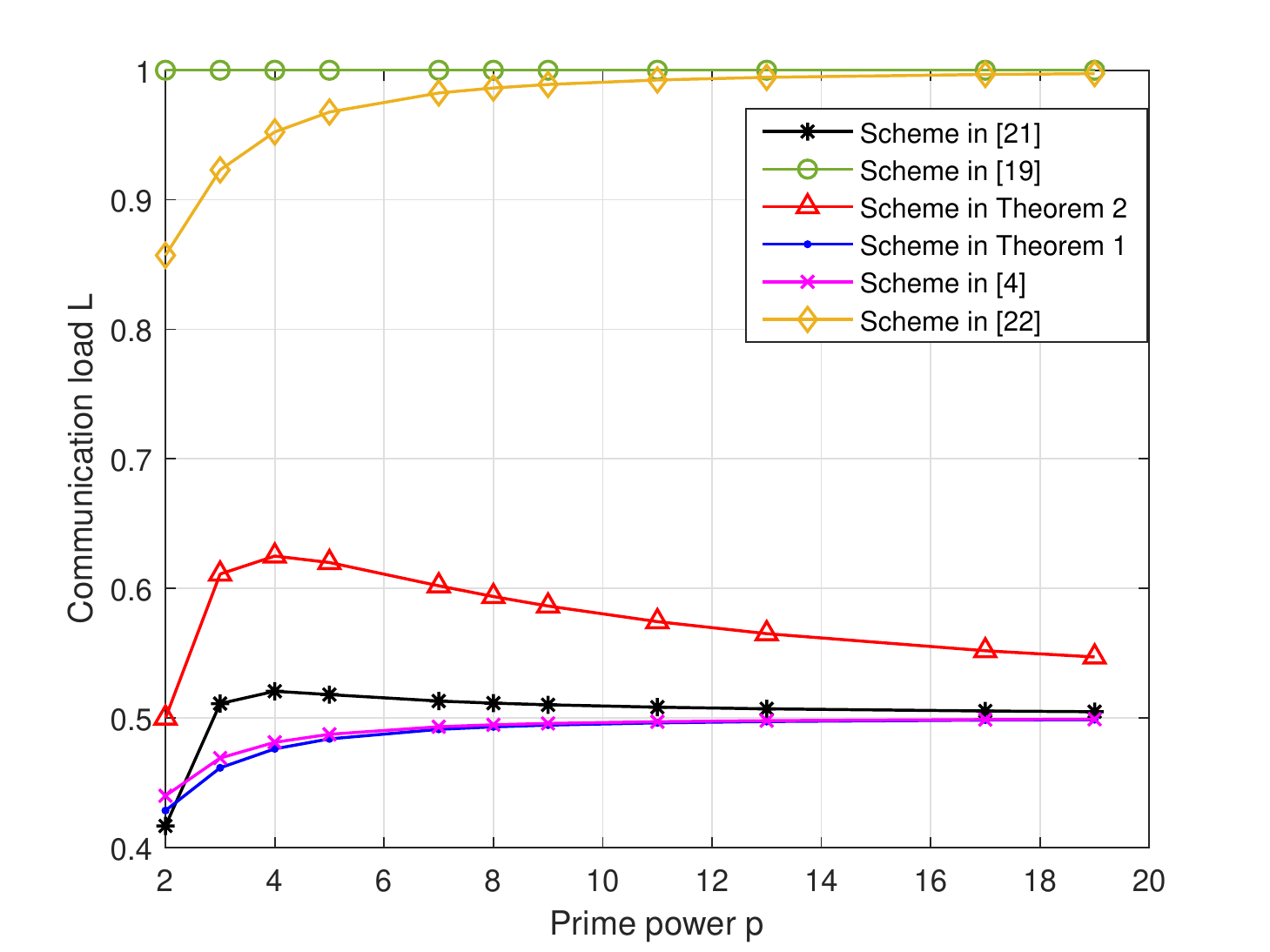}
	\caption{Communication loads of the schemes in Theorems \ref{th-design-CDC} and \ref{th-GDD-CDC} and existing schemes.}
	\label{fig:load}
\end{figure}

Finally, by taking the well-known existing $(p^2+p+1,p+1,1)$ SBIBD and
$2$-$(p,p,p,1)$ GDD for any prime power $p$ \cite{CD}, we obtain the following schemes listed in Table \ref{tab-know-new-speical}.
{
\begin{table*}
  \centering
  \renewcommand\arraystretch{1}
    \setlength{\tabcolsep}{1mm}{
  \caption{Cascaded CDC schemes via $(p^2+p+1,p+1,1)$ SBIBD and $2$-$(p,p,p,1)$ GDD with any prime power $p$. \label{tab-know-new-speical}}
  \begin{tabular}{|c|c|c|c|c|c|c|c|}
\hline
Schemes &\tabincell{c}{Worker number\\  $K$} & \tabincell{c}{Computation \\  Load $r$}& \tabincell{c}{Replication \\ Factor $s$}&  \tabincell{c}{Number of \\ Files $N$}   & \tabincell{c}{Number of Reduce \\ Functions $Q$}&  \tabincell{c}{Communication \\ Load $L_{\text{comm.}}$}& \tabincell{c}{Operation \\ Field $\mathbb{F}_2$}\\
\hline
\cite{WCJ}&\multirow{3}{*}{$p^2$}& \multirow{3}{*}{$p$}&\multirow{3}{*}{$p$}&  \multirow{2}{*}{$p^{p-1}$ } &$p^{p-1}$&
\tabincell{c}{$\frac{1}{2}-\frac{1}{2}\cdot \left(\frac{1}{p}\right)^{p} $\\
$+\left(1-\frac{1}{p}\right)^p \cdot \frac{1}{4p-2}$}&No\\  \cline{1-1}
\cline{6-8}
\cite{JQ} & & & & & $p$&$1$ &Yes\\  \cline{1-1}
\cline{5-8}
Theorem \ref{th-GDD-CDC} && & & $p^2$ &$p^2$& $\frac{1}{2}+\frac{p-2}{p^2}$ &Yes\\[0.3cm]\hline

Theorem \ref{th-design-CDC} &\multirow{2}{*}{\tabincell{c}{\\ $p^2+p+1$}}&\multirow{2}{*}{\tabincell{c}{\\  $p+1$}} &\multirow{2}{*}{\tabincell{c}{\\ $p+1$}} & \multirow{2}{*}{\tabincell{c}{\\ $p^2+p+1$}} &\multirow{2}{*}{\tabincell{c}{\\ $p^2+p+1$}}& $\frac{1}{2}-\frac{1}{2(p^2+p+1)}$ &Yes\\[0.3cm]
\cline{1-1}\cline{7-8}

\cite{JWZ}&& & && &$\frac{p^2+p}{p^2+p+1}$  &No\\
\hline

\end{tabular}
}
\end{table*}
}
Fig. \ref{fig:N}, Fig. \ref{fig:Q}, and Fig. \ref{fig:load} evaluate the number of input files $N$, the number of output functions $Q$, and the communication load versus the prime power $p$ for various schemes, including our schemes and the schemes in \cite{LMYA,WCJ,JQ,JWZ}. Note that Fig. \ref{fig:N} and Fig. \ref{fig:Q} do not plot the scheme in \cite{LMYA} because its parameters $N$ and $Q$   are too large compared to other schemes. 

%\begin{remark}\rm
%\label{remark-performance}

From Table \ref{tab-know-new-speical}, Fig. \ref{fig:N}, Fig. \ref{fig:Q}, and Fig. \ref{fig:load}, the following statements hold.
\begin{itemize}
\item Our scheme for Theorem \ref{th-GDD-CDC} (the red lines in Fig. \ref{fig:N}, Fig. \ref{fig:Q}, and Fig. \ref{fig:load}) achieves almost the same communication load as the scheme in \cite{WCJ} (the black lines in Fig. \ref{fig:N}, Fig. \ref{fig:Q}, and Fig. \ref{fig:load}), while having much smaller file number and output function number when $p$ is large. Compared to the scheme in \cite{JQ} (the black line in Fig. \ref{fig:N} and the green line in Fig. \ref{fig:load}), our scheme has a much smaller file number, slightly larger output function number and much smaller communication load. Compared to   \cite{JWZ} ({the orange line in Fig. \ref{fig:load}}), our scheme has almost the same $N$ and $Q$, but a much smaller communication load.
\item For any positive integer $m$ and any prime power $p\geq 3$, the communication load of our scheme for Theorem \ref{th-design-CDC} (i.e., the blue lines in Fig. \ref{fig:N}, Fig. \ref{fig:Q}, and Fig. \ref{fig:load}) is much less than that of the scheme in \cite{JWZ} with the same order of parameters $K$, $N$, $Q$, $r=s$,   and is smaller than that of the scheme in \cite{LMYA} (i.e., the pink lines in Fig. \ref{fig:load}).
\end{itemize}	
	
%\hfill $\square$
%\end{remark}

%the communication load of the $t$-GDD scheme is larger than that of the $t$-design scheme. However, by Remark \ref{remark-performance-GDD-t=2} and \ref{remark-performance} $t$-GDD scheme has some advantages on the file number, output function number or communication load compared with the existing schemes. %In addition, by the second statement of Remark \ref{remark-relation}, $t$-GDD scheme can achieve some parameters $K$ and $N$ which $t$-design scheme can not achieve.}

\section{Novel Schemes for  Theorem  \ref{th-design-CDC} and Theorem \ref{th-GDD-CDC}}\label{Sec:Schemes}
%In this section, %we introduce how to use combinatorial design structures to construct CDC schemes to improve communication efficiency with small numbers of output functions and input files. The main idea is that
Let the output function arranged set equal to the file stored set defined in \eqref{eq-families}, i.e., $\mathfrak{A}=\mathfrak{D}$, by using  $t$-design (see Definition \ref{def-design}) and $t$-GDD (see Definition \ref{def-GDD}), we can obtain two new cascaded coded distributed computing schemes respectively.

%Before describing our schemes,  we first recall
%\eqref{eq-caches} and \eqref{eq-computing-task} which mean that the storing files strategy and arranging output function strategy are determined by the subsets $\mathcal{D}_n$ (i.e., the worker set each of which stores file $w_n$) and $\mathcal{A}_q$ (i.e., the worker set each of which is arranged to compute the output function $\phi_q(w_{1},w_{2},\ldots,w_{N})$). This implies that in order to design a desired scheme, we only need to study the output function arranged set
%\begin{subequations}
%\label{eq-families}
%\begin{align}
% \mathfrak{A}=\{\mathcal{A}_1,\mathcal{A}_2,\ldots,\mathcal{A}_Q\}, \label{eq-family-1}
%\end{align}
% and the file stored set  \begin{align}  \mathfrak{D}=\{\mathcal{D}_1,\mathcal{D}_2,\ldots,\mathcal{D}_N\}.\label{eq-family-2}
%\end{align}
%\end{subequations}
%
%
%Thus, we can study the output function arranged set $\mathfrak{A}$ and the file stored set $\mathfrak{D}$ defined in \eqref{eq-families} to design cascaded CDC schemes with the computation load and the communication load as small as possible.

\subsection{New CDC Scheme in Theorem \ref{th-design-CDC} via $t$-design}
\label{sub-proof-th-Design-CDC}

%In this subsection we will use a $t$-$(N,M,\lambda)$ design  $(\mathcal{X}, \mathfrak{B})$ to construct a $(K=\lambda{N\choose t}/{M\choose t},r= KM/N=\lambda \binom{N-1}{t-1}/\binom{M-1}{t-1},s=r,N,Q=N)$ cascaded CDC scheme by designing the output function arranged set $\mathfrak{A}$ and the file stored set $\mathfrak{D}$ defined in \eqref{eq-families}.

 Given a $t$-$(N,M,K,r,\lambda)$ design $(\mathcal{X}, \mathfrak{B})$, let $\mathcal{X}=\{x_1$, $x_2$, $\ldots,x_N\}$ and  $\mathfrak{B}=\{\mathcal{B}_1,\mathcal{B}_2,\ldots,\mathcal{B}_K\}$.
%%, where $x_n\in[N]$ and $\mathcal{B}_k\in[K]$ for $n\in[N]$ and $k\in[K]$.
Denote the sets of $N$ files and $N$ output functions by $\mathcal{W}=\{w_{x_1},\ldots,w_{x_N}\}$ and $\mathcal{Q}=\{\phi_{x_1},\ldots,\phi_{x_N}\}$, respectively.

Our main construction idea is as follows:  We first obtain the dual design $(\mathcal{V},\mathfrak{R})$ of design  $(\mathcal{X}, \mathfrak{B})$, and then let the worker set $\mathcal{K}$ be the point set $\mathcal{V}$ and both the output function arranged set $\mathfrak{A}$ and the file stored set $\mathfrak{D}$ be the block set $\mathfrak{R}$, i.e.,
\begin{align}\label{eq-construct-design}
&\mathcal{K}=\mathcal{V}=\mathfrak{B}=\{\mathcal{B}_1,\mathcal{B}_2,\ldots,\mathcal{B}_K\},\\\
&\mathfrak{D}=\mathfrak{A}=\mathfrak{R} = \{\mathcal{R}_{x_1},\mathcal{R}_{x_2},\ldots, \mathcal{R}_{x_N}\}\label{eq-placement},
\end{align}
where $\mathcal{R}_x=\{\mathcal{B}\ |\ x\in \mathcal{B}, \mathcal{B}\in \mathfrak{B}\}$ for  $x\in \mathcal{X}$ as mentioned in Definition \ref{def-dual-design}. Equation \eqref{eq-placement} implies that any worker whose index is in $\mathcal{R}_x$  will store file $w_x$ and compute output function $\phi_x$, for $x\in\mathcal{X}$.
 From Lemma \ref{lemm-dual-design-system}, we have $$K=\frac{\lambda{N\choose t}}{{M\choose t}}, ~~r=s= \frac{KM}{N}=\frac{\lambda \binom{N-1}{t-1}}{\binom{M-1}{t-1}}.$$ Finally, we design the transmission strategy using the cross property of the dual design $(\mathcal{V},\mathfrak{R})$ in Lemma \ref{lemm-dual-design-system}. Specifically,   our scheme operates in the following way.

\vskip 0.2cm
$\bullet$ {\bf Map phase:} According to \eqref{eq-construct-design},  we design the file stored set as $ \mathfrak{D}=\mathfrak{R}$, i.e.,
each worker $\mathcal{B}\in \mathfrak{B}$ stores the files
\begin{align}\label{eq-worker-cache-design}
\mathcal{Z}_{\mathcal{B}}=\{w_{x}\ | \  \mathcal{B}\in \mathcal{R}_x, x\in \mathcal{X}\}.
\end{align}
Then each worker $\mathcal{B}\in \mathfrak{B}$  uses Map functions $\{g_{q,n}(\cdot)\}$ to maps each local input file in $\mathcal{Z}_{\mathcal{B}}$ into $Q$ intermediate values. That is, worker $\mathcal{B}$ could compute the intermediate values
\begin{align}\label{eq-block-design-ZJ}
\mathcal{I}_{\mathcal{B}}=\{v_{y,x}=g_{y,x}(w_{x})\ |\
\mathcal{B}\in \mathcal{R}_x,
x,y\in \mathcal{X} \}
\end{align}%, i.e.,
 %\begin{IEEEeqnarray}{rCl}
%\{v_{q,n}| q\in[Q]\} = g_{q,n}(w_{n}),~\text{for} ~n \in
 %\end{IEEEeqnarray}

%By Lemma \ref{lemm-dual-design-system}, i.e., the dual design $(\mathcal{V},\mathfrak{R})$ is $M$-regular, we have $|\mathcal{Z}_{\mathcal{B}}|=M$. So from \eqref{eq-value-b1} the computation load is
%$$r=\frac{\sum_{\mathcal{B}\in\mathfrak{B}}|\mathcal{Z}_{\mathcal{B}}|}{N}=\frac{KM}{N}=\frac{\lambda M{N\choose t}}{N{M\choose t}}.$$
\begin{example}\rm
\label{example-design-SBIBD-1}
Consider the $(N,M,\lambda)=(7,3,1)$ SBIBD $(\mathcal{X},\mathfrak{B})$ in Example \ref{exam-SBIBD}. By Definition \ref{def-dual-design}, we have its dual design $(\mathcal{V},\mathfrak{R})$ in Example \ref{exam-dual-design}. By Lemma \ref{lemm-dual-design-system},
Then
we have
\begin{align*}
K=\frac{\lambda N(N-1)}{M(M-1)}=\frac{7\cdot 6}{3\cdot 2}=7\ \text{and} \
r=\frac{\lambda (N-1)}{M-1}=\frac{6}{2}=3.
\end{align*} %From \eqref{eq-construct-design}, let $\mathfrak{D}=\mathfrak{A}=\mathfrak{R}$.
Denote the $7$ files by $\mathcal{W}= \{w_{1},w_{2},\ldots,w_{7}\}$ and the $7$ functions by $\mathcal{Q}=\{\phi_{1},\phi_{2},\ldots,\phi_{7}\}$. From \eqref{eq-construct-design} and \eqref{eq-placement} we have $7$ workers $\mathcal{K}=\{\mathcal{B}_1, \mathcal{B}_2, \mathcal{B}_3, \mathcal{B}_4, \mathcal{B}_5, \mathcal{B}_6, \mathcal{B}_7 \}$ and file stored set and output function arranged set
\begin{align}
\mathcal{D}_1&=\mathcal{A}_1=\mathcal{R}_1=\{\mathcal{B}_1, \mathcal{B}_5,\mathcal{B}_7\},\nonumber\\
\mathcal{D}_2&=\mathcal{A}_2=\mathcal{R}_2=\{\mathcal{B}_1,
\mathcal{B}_2,\mathcal{B}_6\},\nonumber\\
\mathcal{D}_3&=\mathcal{A}_3=\mathcal{R}_3=\{\mathcal{B}_2,
\mathcal{B}_3,\mathcal{B}_7\}, \nonumber\\
\mathcal{D}_4&=\mathcal{A}_4=\mathcal{R}_4=\{\mathcal{B}_1,
\mathcal{B}_3,\mathcal{B}_4\},\nonumber\\
\mathcal{D}_5&=\mathcal{A}_5=\mathcal{R}_5=\{\mathcal{B}_2,
\mathcal{B}_4,\mathcal{B}_5\}, \nonumber\\ \mathcal{D}_6&=\mathcal{A}_6=\mathcal{R}_6=\{\mathcal{B}_3,
\mathcal{B}_5,\mathcal{B}_6\},\nonumber\\
\mathcal{D}_7&=\mathcal{A}_7=\mathcal{R}_7=\{\mathcal{B}_4,
\mathcal{B}_6,\mathcal{B}_7\} .
\label{eq-dual-blocks-file}
\end{align}From \eqref{eq-worker-cache-design}, all the workers store the following files respectively.
\begin{align*}
&\mathcal{Z}_{\mathcal{B}_1}=\{w_{1},w_2,w_4\}, \ \  \mathcal{Z}_{\mathcal{B}_2}=\{w_{2},w_3,w_5\}, \\
& \mathcal{Z}_{\mathcal{B}_3}=\{w_{3},w_4,w_6\}, \ \
\mathcal{Z}_{\mathcal{B}_4}=\{w_{4},w_5,w_7\}, \\& \mathcal{Z}_{\mathcal{B}_5}=\{w_{1},w_5,w_6\},  \ \ \mathcal{Z}_{\mathcal{B}_6}=\{w_{2},w_6,w_7\},\\
&\mathcal{Z}_{\mathcal{B}_7}=\{w_{1},w_3,w_7\}.
\end{align*}Then, all the workers can locally compute the following intermediate values.
\begin{align}
&\mathcal{I}_{\mathcal{B}_1}=\{v_{q,n} \ | \ q\in [7], n\in \{1,2,4\}\},\nonumber\\
&\mathcal{I}_{\mathcal{B}_2}=\{v_{q,n}\ |\ q\in [7], n\in \{2,3,5\}\},\nonumber\\
&\mathcal{I}_{\mathcal{B}_3}=\{v_{q,n}\ |\ q\in [7], n\in \{3,4,6\}\},\nonumber\\
&\mathcal{I}_{\mathcal{B}_4}=\{v_{q,n}\ |\ q\in [7], n\in \{4,5,7\}\},\nonumber\\
&\mathcal{I}_{\mathcal{B}_5}=\{v_{q,n}\ |\ q\in [7], n\in \{1,5,6\}\},\nonumber\\
&\mathcal{I}_{\mathcal{B}_6}=\{v_{q,n}\ |\ q\in [7], n\in \{2,6,7\}\},\nonumber\\
&\mathcal{I}_{\mathcal{B}_7}=\{v_{q,n}\ |\ q\in [7], n\in \{1,3,7\}\}.\label{eq-computed-IVs}
\end{align}
%So the computation load is $r=\frac{7\times 3}{7}=3$.		
\hfill $\square$
\end{example}

\vskip 0.2cm
$\bullet$ {\bf Shuffle phase:} According to \eqref{eq-placement}, we set the output function arranged set as  $\mathfrak{R}$, i.e.,  each worker $\mathcal{B}\in \mathfrak{B}$ is arranged to compute the output functions
\begin{align}\label{eq-function-arrangement-design}
\mathcal{Q}_{\mathcal{B}}=\{u_y=\phi_y(w_{
x_1},\ldots,w_{x_N})\ |\ \mathcal{B}\in \mathcal{R}_y, y\in \mathcal{X}\}.
\end{align}  Using the stored files $\mathcal{Z}_{\mathcal{B}}$ and arranged functions $\mathcal{Q}_{\mathcal{B}}$, worker $\mathcal{B}$  requires the following intermediate values which it can not locally compute.
\begin{align}\label{eq-requre-block-design-ZJ}
\mathcal{\overline{I}}_{\mathcal{B}}=\{v_{y,x}=g_{y,x}(w_{x})\ |\
\mathcal{B}\not\in \mathcal{R}_x,\mathcal{B}\in \mathcal{R}_y,
x,y\in \mathcal{X} \}.
\end{align}
From \eqref{eq-block-design-ZJ} and \eqref{eq-requre-block-design-ZJ},  we have the following result that is proved in Appendix \ref{Appendix-IV-class-1}.
\begin{proposition}\rm
\label{pro-required-IVs}
For any two points $x,y\in \mathcal{X}$, $v_{x,y}$ is required and is not locally computed by some worker if and only if $x\neq y$.
\end{proposition}
%\begin{proof}
%Let us consider the first statement, i.e., $x=y$. Assume that the intermediate value $v_{x,x}$ is required by a worker $\mathcal{B}\in \mathcal{V}$. From \eqref{eq-function-arrangement-design} we have $ \mathcal{B}\in \mathcal{R}_x$. Clearly worker $\mathcal{B}$ can locally compute it from \eqref{eq-block-design-ZJ}.
%
%Now let us consider the second statement, i.e., $x \neq y$. Suppose that all the workers require the intermediate value $v_{y,x}$ and can locally computed by themselves. From \eqref{eq-function-arrangement-design} and \eqref{eq-block-design-ZJ}, each worker $\mathcal{B}\in \mathcal{R}_y$ then $\mathcal{B}\in \mathcal{R}_x$, i.e., $\mathcal{R}_y\subseteq \mathcal{R}_x$. Since each block of $\mathfrak{R}$ is $r=\lambda{N-1\choose t-1}/ \binom{M-1}{t-1}$ by the second statement of Lemma \ref{lemm-dual-design-system}. This implies that $|\mathcal{R}_y|=|\mathcal{R}_x|=r$. So we only need to consider the case $\mathcal{R}_y=\mathcal{R}_x$.
%By the third statement of Lemma \ref{lemm-dual-design-system} we have $\lambda_2=|\mathcal{R}_x\bigcap\mathcal{R}_y|=\lambda{N-2\choose t-2}/ \binom{M-2}{t-2}$. Then we have $r=\lambda_2$ which implies $N=M$, a contradiction to Definition \ref{def-design}. Similarly we can also show the intermediate value $v_{x,y}$.
%
%\end{proof}
\begin{example}\rm
\label{example-design-SBIBD-2}
Let us continue using the parameters in Example \ref{example-design-SBIBD-1} to verify the statement in Proposition \ref{pro-required-IVs}.
In the Shuffle phase, from \eqref{eq-dual-blocks-file} and \eqref{eq-computing-task} the functions are arranged for the workers as follows.
\begin{align}
&\mathcal{Q}_{\mathcal{B}_1}=\{\phi_{1},\phi_2,\phi_4\},\ \  \mathcal{Q}_{\mathcal{B}_2}=\{\phi_{2},\phi_3,\phi_5\}, \nonumber\\
&\mathcal{Q}_{\mathcal{B}_3}=\{\phi_{3},\phi_4,\phi_6\}, \ \
\mathcal{Q}_{\mathcal{B}_4}=\{\phi_{4},\phi_5,\phi_7\},\nonumber\\
&\mathcal{Q}_{\mathcal{B}_5}=\{\phi_{1},\phi_5,\phi_6\},  \ \ \mathcal{Q}_{\mathcal{B}_6}=\{\phi_{2},\phi_6,\phi_7\},\nonumber\\
&\mathcal{Q}_{\mathcal{B}_7}=\{\phi_{1},\phi_3,\phi_7\}.
\label{eq-Reduce-Function-arrangement}
\end{align}
Here each output function is computed by $s=3$ workers. From \eqref{eq-computed-IVs} and \eqref{eq-Reduce-Function-arrangement}, the intermediate values required by the workers are listed in Table \ref{tab-BIBD-intermediate-values}.
\begin{table}[!htbp]
\center
\caption{Intermediate values $v_{q,n}$ where $q\in [7]$ and $n\in [7]$ required by workers in $\mathfrak{B}$.
\label{tab-BIBD-intermediate-values}}
\begin{tabular}{c|ccccccc|}
Parameters &\multicolumn{7}{|c|}{worker set $\mathfrak{B}$} \\ \hline
$q,n$&$\mathcal{B}_1$&$\mathcal{B}_2$&$\mathcal{B}_3$&$\mathcal{B}_4$&$\mathcal{B}_5$&$\mathcal{B}_6$&$\mathcal{B}_7$\\ \hline
$q$&
\tabincell{c}{$1,2$,\\ $4$}&\tabincell{c}{$2,3$,\\ $5$}&\tabincell{c}{$3,4$,\\ $6$}&\tabincell{c}{$4,5$,\\ $7$}&\tabincell{c}{$1,5$,\\ $6$}&\tabincell{c}{$2,6$,\\ $7$}&\tabincell{c}{$1,3$,\\ $7$}\\ \hline
$n$&\tabincell{c}{$3,5$,\\ $6,7$}&\tabincell{c}{$1,4$, \\ $6,7$}&\tabincell{c}{$1,2$,\\ $5,7$}&\tabincell{c}{$1,2$,\\ $3,6$}&\tabincell{c}{$2,3$,\\ $4,7$}&
\tabincell{c}{$1,3$,\\ $4,5$}&\tabincell{c}{$2,4$,\\ $5,6$}\\ \hline
\end{tabular}
%}
\end{table}In Table \ref{tab-BIBD-intermediate-values} the first subscript and the second  subscript of each required intermediate value are different.\hfill $\square$
\end{example}

By Proposition \ref{pro-required-IVs}, we only need to consider the intermediate value $v_{x,y}$ where $x,y\in \mathcal{X}$ and $x\neq y$. Using the cross property of dual design, i.e., the third statement of Lemma \ref{lemm-dual-design-system}, we can design our transmission strategy as follows.

\begin{delivery}\rm
\label{delivery-1} For any two distinct points $x,y\in \mathcal{X}$, we randomly choose one worker $\mathcal{B}\in \mathcal{R}_x\bigcap \mathcal{R}_y$ and let it transmit the coded signal $v_{x,y}\oplus v_{y,x}$.
\hfill $\square$
\end{delivery}

Delivery Strategy \ref{delivery-1} always works for the following reasons. By the third statement of Lemma \ref{lemm-dual-design-system}, i.e., any two distinct blocks intersect in exactly
$\lambda_2= \lambda{N-2\choose t-2}/\binom{M-2}{t-2}$
points, we have $\mathcal{R}_x\bigcap \mathcal{R}_y\neq \emptyset$. This implies that we can always choose a worker $\mathcal{B}\in \mathcal{R}_x\bigcap \mathcal{R}_y$. From \eqref{eq-block-design-ZJ}, worker $\mathcal{B}$ can locally compute the intermediate values $v_{x,y}$ and $v_{y,x}$. So worker $\mathcal{B}$ can transmit the coded signal $v_{x,y}\oplus v_{y,x}$.

\vskip 0.2cm
$\bullet$ {\bf Reduce phase:} Now we show that any worker can decode its required intermediate values. Assume that worker $\mathcal{B}_1$ requires the intermediate value $v_{y,x}$ and can not locally compute it. Then from \eqref{eq-requre-block-design-ZJ}, we have $\mathcal{B}_1\not\in \mathcal{R}_x$ and $\mathcal{B}_1\in \mathcal{R}_y$. From \eqref{eq-block-design-ZJ}, worker $\mathcal{B}_1$ can locally compute the intermediate value $v_{x,y}$. So it can decode its required $v_{y,x}$ based on the received coded signal $v_{x,y}\oplus v_{y,x}$.

Clearly, there are $\binom{N}{2}$ coded signals that are transmitted in Delivery Strategy \ref{delivery-1}. Recall that each intermediate value has $T$ bits.  Then the communication load is $
L_{t\text{-design}}=\frac{\binom{N}{2}\cdot T}{NQT}=\frac{N(N-1)}{2N^2}=\frac{N-1}{2N}$.
Thus, using $t$-design, we can obtain Theorem \ref{th-design-CDC}.

\begin{example}\rm
\label{example-design-SBIBD-3}
Let us continue using the parameters in Examples \ref{example-design-SBIBD-1} and \ref{example-design-SBIBD-2}. Let us first consider the pair of intermediate values $v_{1,2}$ and $v_{2,1}$ in Table \ref{tab-BIBD-intermediate-values}. We have $ \mathcal{B}_1 =\mathcal{D}_{1}\bigcap \mathcal{D}_{2}$ from \eqref{eq-dual-blocks-file}. From \eqref{eq-computed-IVs} worker $ \mathcal{B}_1 $ can compute intermediate values $v_{1,2}$ and $v_{2,1}$. Then worker $\mathcal{B}_1$ transmits $v_{1,2}\oplus v_{2,1}$. By Table \ref{tab-BIBD-intermediate-values}, $v_{1,2}$ is required by workers $\mathcal{B}_5$, $\mathcal{B}_7$, and  $v_{2,1}$ is required by workers $\mathcal{B}_2$, $\mathcal{B}_6$. So in Reduce phase, after receiving $v_{1,2}\oplus v_{2,1}$, worker $\mathcal{B}_2$ and worker $\mathcal{B}_6$ can individually decode the requiring intermediate value $v_{2,1}$ with the locally computed intermediate value $v_{1,2}$; worker $\mathcal{B}_5$ and worker $\mathcal{B}_7$ can individually decode the requiring intermediate value $v_{1,2}$ with its locally computed intermediate value $v_{2,1}$. Similarly,
all the workers can send the coded signals listed in Table \ref{tab-BIBD-delivery} and each worker $\mathcal{B}\in \mathcal{K}$ can compute each output function in $\mathcal{Q}_{\mathcal{B}}$. Since there are $21$ transmitted signals in  Table \ref{tab-BIBD-delivery}, the communication load is $L_{t\text{-design}}=\frac{21}{7^2}=\frac{3}{7}$.

 When $K=7$, $r=s=3$, we can obtain a cascaded CDC scheme in \cite{LMYA} with $N=Q=35$ and communication load $L_{\text{comm.}}=\frac{11}{25}$, which is larger than the communication load of ours\footnote{\label{compare-li} The communication load in \cite{LMYA} is optimal under a  specific output function assignment satisfying $ \mathfrak{A}={[K]\choose s}$. Although different schemes  may have different values $Q$ and $N$, the communication load defined in \eqref{eq-communication-load} is normalized by $NQT$, implying that if $(K,r,s)$ are consistent in different schemes, it is fair to compare their communication loads with different $(N,Q)$.  } and a scheme in \cite{JWZ} with $N=Q=7$ and communication load $L_{\text{comm.}}=\frac{6}{7}$ which is larger than the communication of ours.
		\hfill $\square$

\begin{table}
  \centering
\caption{Coded signals sent by workers in $\mathfrak{B}$.
\label{tab-BIBD-delivery}}
  \begin{tabular}{c|ccc}
    \hline
$\mathcal{B}_1$&
     $v_{1,2}\oplus v_{2,1}$ & $v_{1,4}\oplus v_{4,1}$ & $v_{2,4}\oplus v_{4,2}$ \\
$\mathcal{B}_2$&
     $v_{2,3}\oplus v_{3,2}$ & $v_{2,5}\oplus v_{5,2}$ & $v_{3,5}\oplus v_{5,3}$ \\
$\mathcal{B}_3$&
    $v_{3,4}\oplus v_{4,3}$&$v_{3,6}\oplus v_{6,3}$&$v_{4,6}\oplus v_{6,4}$\\
$\mathcal{B}_5$&
    $v_{4,5}\oplus v_{5,4}$&$v_{4,7}\oplus v_{7,4}$&$v_{5,7}\oplus v_{7,5}$\\
$\mathcal{B}_6$&
    $v_{1,5}\oplus v_{5,1}$&$v_{1,6}\oplus v_{6,1}$&$v_{5,6}\oplus v_{6,5}$\\
$\mathcal{B}_7$&
    $v_{2,6}\oplus v_{6,2}$&$v_{2,7}\oplus v_{7,2}$&$v_{6,7}\oplus v_{7,6}$\\
$\mathcal{B}_7$&
    $v_{1,3}\oplus v_{3,1}$&$v_{1,7}\oplus v_{7,1}$&$v_{3,7}\oplus v_{7,3}$\\ \hline
  \end{tabular}
\end{table}
\end{example}
\begin{remark}[Multicast gain]\rm
\label{remark-coded-gain}
In Delivery Strategy \ref{delivery-1}, each of the intermediate values $v_{y,x}$ and $v_{x,y}$ is required by
\begin{align*}
|\mathcal{R}_x\setminus \mathcal{R}_y|=|\mathcal{R}_y\setminus \mathcal{R}_x|=s-\lambda_2=r-\lambda_2
\end{align*} workers who can not locally compute it themselves. So the coded signal $v_{y,x}\oplus v_{x,y}$ is useful for $2(r-\lambda_2)$ workers, i.e., the multicast gain is $2(r-\lambda_2)$.
%So our scheme generated by a $(N,M,K,r,1)$ BIBD achieves a multicast gain $2(r-\lambda_2)=2r-2$ that is just one less than the maximum multicast gain of the scheme proposed in \cite{LMYA}.
\end{remark}

\subsection{New CDC Scheme in Theorem \ref{th-GDD-CDC} via  $t$-GDD}
\label{sub-proof-th-GDD-CDC}
In Delivery Strategy \ref{delivery-1}, the key point for designing the delivery strategy is the cross property of dual design. That is, for any design $(\mathcal{X},\mathfrak{R})$, if its dual design $(\mathcal{V}, \mathfrak{R})$ satisfies that $\mathcal{R}_x\bigcap \mathcal{R}_y\neq \emptyset$ for any two different points $x$, $y\in \mathcal{X}$, the Delivery Strategy \ref{delivery-1} also works.

If there exist two different points $x$, $y\in \mathcal{X}$ such that $\mathcal{R}_x\bigcap\mathcal{R}_y= \emptyset$, Delivery Strategy \ref{delivery-1} can be modified by just transmitting the intermediate values $v_{y,x}$ and $v_{x,y}$ respectively. Specifically we can randomly choose two workers, say $\mathcal{B}_1\in \mathcal{R}_x$ and  $\mathcal{B}_2\in \mathcal{R}_y$, and then let workers $\mathcal{B}_1$ and $\mathcal{B}_2$ transmit intermediate values $v_{y,x}$ and $v_{x,y}$ respectively.

To verify our claim, we first take the $t$-$(m ,q, M,\lambda)=2$-$(3,2,3,1)$ GDD $(\mathcal{X}, \mathfrak{G}, \mathfrak{B})$ in Example \ref{exam-1-t-GDD} to generate a $(K,r,s,N,Q=N)=(4,2,2,6,6)$ cascaded CDC scheme.

\begin{example}\rm
\label{exam-t-GDD-CDC}
By Example \ref{exam-t-dual-GDD}, we have the dual design $(\mathcal{V},\mathfrak{R})$  of the $(\mathcal{X}, \mathfrak{G}, \mathfrak{B})$ in Example \ref{exam-1-t-GDD}. Denote $K=4$ workers, $6$ files and functions by  $\mathcal{K}=\mathcal{V}=\mathfrak{B}=\{\mathcal{B}_1,\mathcal{B}_2,
\mathcal{B}_3,\mathcal{B}_4\}$, $\mathcal{W}=\{w_{1}, w_{2}$, $\ldots,w_{6}\}$ and  $\mathcal{Q}=\{\phi_{1},\phi_{2},\ldots,\phi_{6}\}$ respectively. Now we introduce the three phases of cascaded CDC scheme as follows.
%\begin{itemize}
%\item

$\bullet$ {\bf Map phase:} Let $\mathfrak{D}=\mathfrak{A}=\mathfrak{R}$. Then from \eqref{eq-dual-GDD-blocks} we have
\begin{align}
\mathcal{D}_1&=\mathcal{A}_1=\mathcal{R}_1=\{\mathcal{B}_1,\mathcal{B}_2\},\
\mathcal{D}_2 =\mathcal{A}_2=\mathcal{R}_2=\{\mathcal{B}_3,\mathcal{B}_4\},\nonumber\\
\mathcal{D}_3&=\mathcal{A}_3=\mathcal{R}_3=\{\mathcal{B}_1,\mathcal{B}_4\}, \
\mathcal{D}_4=\mathcal{A}_4=\mathcal{R}_4=\{\mathcal{B}_2,\mathcal{B}_3\},\nonumber\\ \mathcal{D}_5&=\mathcal{A}_5=\mathcal{R}_5=\{\mathcal{B}_1,\mathcal{B}_3\}, \
\mathcal{D}_6 =\mathcal{A}_6=\mathcal{R}_6=\{\mathcal{B}_2,\mathcal{B}_4\}.
    \label{eq-dual-GDD-blocks-file-function}
\end{align}
From \eqref{eq-caches}, all the workers store the following files respectively.
\begin{align*}
&\mathcal{Z}_{\mathcal{B}_1}=\{w_{1},w_3,w_5\},\ \mathcal{Z}_{\mathcal{B}_2}=\{w_{1},w_4,w_6\},\\
&\mathcal{Z}_{\mathcal{B}_3}=\{w_{2},w_4,w_5\},\ \mathcal{Z}_{\mathcal{B}_4}=\{w_{2},w_3,w_6\}.
\end{align*}Then all the workers can locally compute the following intermediate values.
\begin{align}
&\mathcal{I}_{\mathcal{B}_1}=\{v_{q,n}\ |\ q\in [6], n\in \{1,3,5\}\},\nonumber\\
&\mathcal{I}_{\mathcal{B}_2}=\{v_{q,n}\ |\ q\in [6], n\in \{1,4,6\}\},\nonumber\\
&\mathcal{I}_{\mathcal{B}_3}=\{v_{q,n}\ |\ q\in [6], n\in \{2,4,5\}\},\nonumber\\
&\mathcal{I}_{\mathcal{B}_4}=\{v_{q,n}\ |\ q\in [6], n\in \{2,3,6\}\}.\label{eq-computed-IVs-2}
\end{align}
%So the computation load is $r=\frac{KM}{N}=\frac{4\times 3}{6}=2$.

$\bullet$ {\bf Shuffle phase:} From \eqref{eq-dual-GDD-blocks-file-function} and \eqref{eq-computing-task} the output functions are arranged for workers as follows.
\begin{align}\label{eq-Reduce-Function-arrangement-2}
&\mathcal{Q}_{\mathcal{B}_1}=\{\phi_{1},\phi_3,\phi_5\},\ \  \mathcal{Q}_{\mathcal{B}_2}=\{\phi_{1},\phi_4,\phi_6\}, \nonumber\\
&\mathcal{Q}_{\mathcal{B}_3}=\{\phi_{2},\phi_4,\phi_5\}, \ \
\mathcal{Q}_{\mathcal{B}_4}=\{\phi_{2},\phi_3,\phi_6\}.
\end{align}
 From \eqref{eq-computed-IVs-2} and \eqref{eq-Reduce-Function-arrangement-2}, the intermediate values that are required and can not be computed locally by the workers are listed in Table \ref{tab-BIBD-intermediate-values-1}. We can see that for any two points $x,y\in \mathcal{X}$, the intermediate value $v_{x,y}$ is in Table \ref{tab-BIBD-intermediate-values-1} if and only if $x\neq y$, i.e., the subscripts of each intermediate value also satisfy Proposition \ref{pro-required-IVs}. So similar to Delivery Strategy \ref{delivery-1}, we can design the delivery strategy as follows.
\begin{table}[!htbp]
\center
\caption{Intermediate values $v_{q,n}$ where $q,n\in [6]$ required and can not be locally computed by workers in $\mathfrak{B}$.
\label{tab-BIBD-intermediate-values-1}}
\begin{tabular}{c|cccc|}
Parameters &\multicolumn{4}{|c|}{worker set $\mathfrak{B}$} \\ \hline
$q,n$&$\mathcal{B}_1$&$\mathcal{B}_2$&$\mathcal{B}_3$&$\mathcal{B}_4$\\ \hline
$q$&
$1,3,5$&$1,4,6$&$2,4,5$&$2,3,6$\\ \hline
$n$&$2,4,6$&$2,3,5$&$1,3,6$&$1,4,5$\\ \hline
\end{tabular}
%}
\end{table}
Let us consider the intermediate value $v_{1,3}$. From \eqref{eq-dual-GDD-blocks-file-function} we have $\mathcal{R}_1\bigcap \mathcal{R}_3=\{\mathcal{B}_1\}$. Using Delivery Strategy \ref{delivery-1}, we let worker $\mathcal{B}_1$ transmit the coded signal $v_{1,3}\bigoplus v_{3,1}$. Let us consider another intermediate value $v_{1,2}$ in Table \ref{tab-BIBD-intermediate-values-1}. From \eqref{eq-dual-GDD-blocks-file-function} we have $\mathcal{R}_1\bigcap \mathcal{R}_2=\emptyset$. Then we randomly choose workers $\mathcal{B}_1\in \mathcal{R}_1$ and $\mathcal{B}_3\in \mathcal{R}_2$, then let them transmit intermediate values $v_{1,2}$ and $v_{2,1}$ respectively. Similar to the above delivery strategy, we can obtain all the transmitted signals listed in Table \ref{tab-GDD-delivery} where there are $18$ transmitted signals.

		\begin{table}[!htbp]
\center
\caption{Coded signals sent by workers in $\mathfrak{B}$.
\label{tab-GDD-delivery}}
\begin{tabular}{|cccc|}
\multicolumn{4}{|c|}{worker set $\mathfrak{B}$} \\ \hline
$\mathcal{B}_1$&$\mathcal{B}_2$&$\mathcal{B}_3$&$\mathcal{B}_4$\\ \hline
$v_{1,3}+v_{3,1}$&$v_{1,4}+v_{4,1}$&$v_{2,4}+v_{4,2}$&$v_{2,3}+v_{3,2}$\\
$v_{1,5}+v_{5,1}$&$v_{1,6}+v_{6,1}$&$v_{2,5}+v_{5,2}$&$v_{2,6}+v_{6,2}$\\
$v_{3,5}+v_{5,3}$&$v_{4,6}+v_{6,4}$&$v_{4,5}+v_{5,4}$&$v_{3,6}+v_{6,3}$\\
$v_{2,1}$&$v_{3,4}$&$v_{1,2}$&$v_{4,3}$\\
$v_{6,5}$&&&$v_{5,6}$\\
\hline
\end{tabular}
%}
\end{table}We can check that all the workers can decode their required intermediate values. Then the communication load is $L_{t\text{-GDD}}=\frac{18}{6^2}=\frac{1}{2}$.
\hfill $\square$

\end{example}

According to Example \ref{exam-t-GDD-CDC}, we are ready to present the general construction of a $(K$, $r$, $ s=r$, $N=mq$, $Q=mq)$ cascaded CDC scheme via a $t$-$(m ,q, M,\lambda)$ GDD $(\mathcal{X}, \mathfrak{G}, \mathfrak{B})$ where $K=\lambda{m\choose t}q^t/{M\choose t}$ and $r=\lambda q^{t-1}\binom{m-1}{t-1}/\binom{M-1}{t-1}$.

Let $\mathcal{X}=\{x_1,x_2,\ldots,x_{mq}\}$, $\mathfrak{G}=\{\mathcal{G}_1,\mathcal{G}_2,\ldots,\mathcal{G}_m\}$ and $\mathfrak{B}=\{\mathcal{B}_1,\mathcal{B}_2,\ldots,\mathcal{B}_K\}$. Denote its dual design by $(\mathcal{V},\mathfrak{R})$. Using the same constructing method in \ref{sub-proof-th-Design-CDC}, i.e.,  \eqref{eq-construct-design}, we have the worker set $\mathcal{K}=\mathcal{V}=\mathfrak{B}$ and the file stored set and the output function arranged set $\mathfrak{D}=\mathfrak{A}=\mathfrak{R}$. By Lemma \ref{lemm-dual-GDD-system}, we have $K=\lambda{m\choose t}q^t/{M\choose t}$ and $r=s=\lambda q^{t-1}\binom{m-1}{t-1}/\binom{M-1}{t-1}$.

Denote the $mq$ files and output functions as $\mathcal{W}=\{w_{x_1},w_{x_2},\ldots,w_{x_{mq}}\}$  and  $\mathcal{Q}=\{\phi_{x_1},\phi_{x_2},\ldots,\phi_{x_{mq}}\}$, respectively. In the Map and Shuffle phases, we can also obtain the file stored set and the output function arranged set for each worker $\mathcal{B}\in \mathfrak{B}$ as   listed in \eqref{eq-worker-cache-design} and \eqref{eq-function-arrangement-design} respectively. Using \eqref{eq-worker-cache-design} and \eqref{eq-function-arrangement-design}, the sets of locally computed intermediate values and desired intermediate values (not locally computed but required to compute the assigned output functions) by worker $\mathcal{B}\in \mathfrak{B}$ can be written as the same form as \eqref{eq-block-design-ZJ} and \eqref{eq-requre-block-design-ZJ}, respectively. Then, from \eqref{eq-block-design-ZJ} and \eqref{eq-requre-block-design-ZJ}, we can also show that all the intermediate values also satisfy Proposition \ref{pro-required-IVs}.

So it is sufficient to consider the delivery strategy in the Shuffle phase. By the third and fourth statements of Lemma \ref{lemm-dual-GDD-system}, we have the following delivery strategy.
\begin{delivery}\rm
\label{delivery-2}
For any two distinct points $x,y\in \mathcal{X}$, our delivery strategy consists of the following two cases depending on whether $\mathcal{R}_x\bigcap \mathcal{R}_y$ is empty or not.
\begin{itemize}
\item[1)]If $\mathcal{R}_x\bigcap \mathcal{R}_y=\emptyset$, we randomly choose one worker $\mathcal{B}_1\in \mathcal{R}_x$ and $\mathcal{B}_2\in \mathcal{R}_y$ and let workers $\mathcal{B}_1$ and $\mathcal{B}_2$ transmit the intermediate values $v_{x,y}$ and $v_{y,x}$ respectively.
\item[2)] If $\mathcal{R}_x\bigcap \mathcal{R}_y\neq\emptyset$, we also use the Delivery Strategy \ref{delivery-1}, i.e., we randomly choose a worker $\mathcal{B}\in \mathcal{R}_x\bigcap \mathcal{R}_y$ and let it transmit $v_{x,y}\bigoplus v_{y,x}$.\hfill $\square$
\end{itemize}
\end{delivery}

In Delivery Strategy \ref{delivery-2}, we can always choose workers $\mathcal{B}_1\in \mathcal{R}_x$ and $\mathcal{B}_2\in \mathcal{R}_y$ since each block of $\mathfrak{R}$ has exactly $r$ points by the first statement of Lemma \ref{lemm-dual-GDD-system}. Similar to Subsection \ref{sub-proof-th-Design-CDC}, from \eqref{eq-block-design-ZJ} and \eqref{eq-requre-block-design-ZJ} we can also show that each worker can decode its required but not locally computed intermediate values.

Finally, let us compute the communication load. By the third  statement of Lemma \ref{lemm-dual-GDD-system}, there are exactly $m{q\choose 2}$ pairs $\{x,y\}\subseteq\mathcal{X}$ where $x\neq y$. So there are exactly $2m{q\choose 2}$ transmitted signals in the first case of Delivery Strategy \ref{delivery-2}. By the fourth statement of Lemma \ref{lemm-dual-GDD-system}, there are exactly ${mq\choose 2}-m{q\choose 2}$ pairs $\{x,y\}\subseteq\mathcal{X}$ where $x\neq y$. So there exactly ${mq\choose 2}-m{q\choose 2}$ transmitted signals in the second case of Delivery Strategy \ref{delivery-2}. Recall that each intermediate value has $T$ bits. Then the communication load is
\begin{align}\label{eq-trans-1}
L_{t\text{-GDD}}&=\frac{({mq\choose 2}-m{q\choose 2})T +2m{q\choose 2}T}{NQT}\\
&=\frac{mq(mq-1)+mq(q-1)}{2(mq)^2}=\frac{1}{2}+\frac{q-2}{2mq}. \nonumber
\end{align}

Finally, we compute the average multicast gain.
%g=\frac{\sum\limits_{k=1}^K \sum\limits_{q\in\mathcal{Q}_k}\sum\limits_{n\in [N]\backslash\mathcal{Z}_k} |v_{q,n}|}{\sum\limits_{k=1}^K|X_k|} =\frac{sQ(N-M)T}{\sum\limits_{k=1}^K l_k}.
\begin{remark}\rm
\label{remark-coded-gain-GDD}
From \eqref{eq-worker-cache-design} and by the first statement of Lemma \ref{lemm-dual-GDD-system}, i.e., each point of $\mathcal{V}$ occurs in exactly $M$ blocks and from \eqref{eq-mul-gain}, the multicast gain of the scheme in Theorem \ref{th-GDD-CDC} is
\begin{align*}
g_{t\text{-GDD}}&=\frac{sQ(N-M)T}{\sum\limits_{k=1}^K l_k}
=\frac{rmq(mq-M)T}{\left({mq\choose 2}-m{q\choose 2}\right)\cdot T+mq(q-1)T}
\nonumber\\
&=2r-\frac{2r(M+q-2)}{mq+q-2}.
\end{align*}
\end{remark}

\section{Extentions:  Case   $r\neq s$}
\label{sec-extension}
 In this section, we   show how our constructing method can be extended to the case $r\neq s$. We will use a $t$-$(N,M,\lambda)$ design  $(\mathcal{X}, \mathfrak{B})$ to construct a $(K=\lambda{N\choose t}/{M\choose t},r =\lambda{N-1\choose t-1}/ \binom{M-1}{t-1},s=\frac{\lambda(N-t+1)}{ M-t+1},N,Q={N\choose t-1})$ cascaded CDC scheme by designing the output function arranged set $\mathfrak{A}$ and the file stored set $\mathfrak{D}$ defined in \eqref{eq-families}.

Similar to the construction of $t$-design scheme, we   first obtain the dual design $(\mathcal{V},\mathfrak{R})$ of design $(\mathcal{X}, \mathfrak{B})$, and let $\mathcal{K}=\mathcal{V}$ and $\mathcal{D}=\mathfrak{R}$. Then we have $K=\lambda{N\choose t}/{M\choose t}$ and $r=\lambda{N-1\choose t-1}/ \binom{M-1}{t-1}$ by Lemma \ref{lemm-dual-design-system}. Different from the definition of the output function arranged set in \eqref{eq-placement}, here we choose the output function arranged set as
\begin{align}
\label{eq-function-r-neq-s}
\mathfrak{A}=\left\{\mathcal{A}=\bigcap\limits_{x\in \mathcal{C}}\mathcal{R}_x\ \Big|\ \mathcal{C}\in{\mathcal{X}\choose t-1}\right\}.
\end{align} Clearly $Q={N\choose t-1}$. By the third statement of Lemma \ref{lemm-dual-design-system}, we have
$s=\lambda_{t-1}=\frac{\lambda(N-t+1)}{M-t+1}$.

Let $\mathcal{X}=\{x_1,x_2,\ldots,x_N\}$ and $\mathfrak{B}=\{\mathcal{B}_1,\mathcal{B}_2,\ldots,\mathcal{B}_K\}$. Denote $N$ files by $\mathcal{W}=\{w_{x_1},w_{x_2},\ldots,w_{x_N}\}$ and $Q={N\choose t-1}$ functions $\mathcal{Q}=\{\phi_{\mathcal{A}_1},\phi_{\mathcal{A}_2},
\ldots,\phi_{\mathcal{A}_Q}\}$. We can obtain our desired scheme in the following way.

In the Map phase, let $\mathfrak{D}=\mathfrak{R}$. Then each worker $\mathcal{B}\in \mathfrak{B}$ stores the file set $\mathcal{Z}_{\mathcal{B}}$ which is listed in \eqref{eq-worker-cache-design} since the file stored set are the same as that of $t$-design scheme, and could compute the intermediate values
\begin{align}
\label{eq-r-neq-s-computed}
\mathcal{I}_{\mathcal{B}}=\left\{v_{\mathcal{A},x}
=g_{\mathcal{A},x}(w_{x})\ \Big| \ \mathcal{B}\in \mathcal{R}_x,x\in \mathcal{X} \right\}.
\end{align}

In the shuffle phase, from \eqref{eq-function-r-neq-s} and \eqref{eq-computing-task}, each worker $\mathcal{B}\in \mathfrak{B}$ is arranged to compute the output functions
\begin{align}
\label{eq-r-neq-s-arranged-function}
\mathcal{Q}_{\mathcal{B}}=\left\{u_{\mathcal{A}}=\phi_{\mathcal{A}}(w_{x_1},w_{x_2},\ldots,w_{x_N})\ \Big| \mathcal{B}\in \mathcal{A},\mathcal{A}\in \mathfrak{A} \right\}.
\end{align}
From \eqref{eq-r-neq-s-computed} and \eqref{eq-r-neq-s-arranged-function}, each worker $\mathcal{B}$ requires the following intermediate values
\begin{align}\label{eq-r-neq-s-required}
\mathcal{\overline{I}}_{\mathcal{B}}=
\{v_{\mathcal{A},x}\ |\ \mathcal{B}\in \mathcal{A}, \mathcal{A}\in\mathfrak{A}, \mathcal{B}\not\in \mathcal{R}_x,x\in \mathcal{X}
\}.
\end{align}
From \eqref{eq-r-neq-s-computed},  \eqref{eq-function-r-neq-s}, and \eqref{eq-r-neq-s-required}, the following result can be obtained.
\begin{proposition}\rm
\label{pro-propery-r-neq-s}
For any $x\in \mathcal{X}$, $\mathcal{C}\in {\mathcal{X}\choose t-1}$ and any block $\mathcal{B}\in \mathfrak{B}$, let $\mathcal{A}=\bigcap_{x\in \mathcal{C}}\mathcal{R}_x$. Then we have that
\begin{itemize}
\item the intermediate value $v_{\mathcal{A},x}$ is required and is not locally computed by worker $\mathcal{B}$ if and only if $\mathcal{B}\in \bigcap_{x\in \mathcal{C}}\mathcal{R}_x$,  $\mathcal{A}\in\mathfrak{A}$,  and $\mathcal{B}\not\in \mathcal{R}_x$, and
\item the intermediate value $v_{\mathcal{A},x}$ is locally computed by worker $\mathcal{B}$ if and only if $ \mathcal{B}\in \mathcal{R}_x$.\hfill $\square$
\end{itemize}
\end{proposition}
By Proposition \ref{pro-propery-r-neq-s}, we design the following delivery strategy to exchange the intermediate values among the workers.

\begin{delivery}\rm
\label{delivery-3} For any $t$-subset $\mathcal{T}\in {\mathcal{X}\choose t}$, we randomly choose a worker $\mathcal{B}\in \bigcap_{x\in \mathcal{T}}\mathcal{R}_x$, and let it transmit the coded signal
\begin{align}\label{eq-coded-sginal-r-neq-s}
X_{\mathcal{T}}=\bigoplus\limits_{\mathcal{C}\in {\mathcal{T}\choose t-1},\mathcal{A}=\bigcap\limits_{x\in \mathcal{C}}\mathcal{R}_x}v_{\mathcal{A},\mathcal{T}\setminus \mathcal{C}}.
\end{align}
\vskip -0.8cm \hfill $\square$
\end{delivery}
In Delivery Strategy \ref{delivery-3}, we can always choose a worker $\mathcal{B}\in \mathcal{B}\in \bigcap_{x\in \mathcal{T}}\mathcal{R}_x$ by the third statement of Lemma \ref{lemm-dual-design-system}, i.e., the intersection of any $t$ blocks of $\mathfrak{R}$ has $\lambda>0$ points. Then for each $x\in \mathcal{T}$ we have $\mathcal{B}\in \mathcal{R}_x$. From \eqref{eq-r-neq-s-computed} worker $\mathcal{B}$ can compute all the intermediate values $v_{\mathcal{A},\mathcal{T}\setminus\mathcal{C}}$ in \eqref{eq-coded-sginal-r-neq-s}.

Now let us show that each worker requires one intermediate value of $X_{\mathcal{T}}$ in the Reduce phase. Assume that a worker $\mathcal{B}'$ requires a intermediate value $v_{\mathcal{A}',\mathcal{T}\setminus\mathcal{C}'}$. By the first statement of Proposition \ref{pro-propery-r-neq-s}, we have $
\mathcal{B}'\in \bigcap_{x\in \mathcal{C}'}\mathcal{R}_x$, which implies that worker $\mathcal{B}'$ can locally compute the other intermediate values by the second statement of Proposition \ref{pro-propery-r-neq-s}.

 By Delivery Strategy \ref{delivery-3}, there are ${N\choose t}$ transmitted coded signals. Recall that each intermediate value has $T$ bits. Thus, we achieve the communication load $L_{\text{comm.}}=\frac{{N\choose t}T}{N\binom{N}{t-1}T}=\frac{N-t+1}{Nt}$.

 We formally state this result in the following Theorem.
\begin{theorem}\rm
\label{th-design-CDC-rs}
If there exists a $t$-$(N,M,\lambda)$ design, we can obtain a $(K=\lambda{N\choose t}/{M\choose t} $, $M$, $r=\lambda {N-1\choose t-1}/{M-1\choose t-1}$, $s= \frac{\lambda(N-t+1)}{ M-t+1}$, $N$, $Q={N\choose t-1})$ cascaded CDC scheme with  the communication load $L_{\text{comm.}}=\frac{N-t+1}{Nt}$.
		\hfill $\square$
\end{theorem}

Finally, let us take  a $3$-$(N_1,M_1,1)$ design to illustrate the superiority of our scheme in Theorem \ref{th-design-CDC-rs} over the state-of-art works. Since \cite{WCJ} and \cite{JWZ}  both focus on the case with $r=s$,   we only need to compare with the scheme in \cite{JQ}. By Theorem \ref{th-design-CDC-rs}, the scheme in the second row of Table \ref{tab-Main-results-t=3-r-s} is obtained.

By Table \ref{tab-known-CDC}, we can obtain a scheme in \cite{JQ} with $K=\frac{N_1(N_1-1)(N_1-2)}{M_1(M_1-1)(M_1-2)}$ distributed computing workers, $N=(\frac{N_1}{M_1})^{\frac{(N_1-1)(N_1-2)}{(M_1-1)(M_1-2)}-1}$ files, $Q=\frac{N_1(N_1-1)}{M_1(M_1-1)}$ output functions such that each output function is computed by $s= \frac{N_1-2}{ M_1-2}$ workers, the computation load $r=\frac{(N_1-1)(N_1-2)}{(M_1-1)(M_1-2)}$, and the communication load
\begin{align*}
L_{\text{comm.}}&= \frac{s}{r-1}\cdot\left(1-\frac{r}{K}\right)\\
&=\frac{ \frac{N_1-2}{ M_1-2}}{\frac{(N_1-1)(N_1-2)}{(M_1-1)(M_1-2)}-1}\cdot
\left(1-\frac{\frac{(N_1-1)(N_1-2)}{(M_1-1)(M_1-2)}}{\frac{N_1(N_1-1)(N_1-2)}{M_1(M_1-1)(M_1-2)}}\right)\\
&=\frac{1}{\frac{N_1-1}{M_1-1}-\frac{ M_1-2}{N_1-2}}\cdot
\left(1-\frac{M_1}{N_1}\right)\\
&<\frac{1}{\frac{N_1-1}{M_1-1}-\frac{ M_1-1}{N_1-1}}\cdot
\left(1-\frac{M_1}{N_1}\right)\\
&=\frac{N_1-1}{N_1}\cdot\frac{M_1-1}{N_1+M_1}.
\end{align*}  From Table \ref{tab-Main-results-t=3-r-s}, we can see that under the same parameters $K$, $r$ and $s$, the scheme in Theorem \ref{th-design-CDC-rs} has a much smaller file number $N$ while increasing some communication load compared with the scheme in \cite{JQ}.
{\begin{table*}
  \centering
  \renewcommand\arraystretch{1}
    \setlength{\tabcolsep}{1mm}{
  \caption{The schemes in \cite{LMYA,WCJ,JWZ,JQ} and the scheme via $3$-$(N_1,M_1,1)$ design in Theorem \ref{th-design-CDC-rs}\label{tab-Main-results-t=3-r-s} }
  \begin{tabular}{|c|c|c|c|c|c|c|c|}
\hline
Parameters &\tabincell{c}{Worker number\\  $K$} & \tabincell{c}{Computation \\  Load $r$}& \tabincell{c}{Replication \\ Factor $s$}&  \tabincell{c}{Number of \\ Files $N$}   & \tabincell{c}{Number of Reduce \\ Functions $Q$}&  \tabincell{c}{Communication \\ Load $L_{\text{comm.}}$}& \tabincell{c}{Operation \\ Field $\mathbb{F}_2$}\\
\hline

Theorem \ref{th-design-CDC-rs} &
\multirow{2}{*}{$\frac{N_1 (N_1 -1)(N_1 -2)}{M_1 (M_1 -1)(M_1 -2)}$} & \multirow{2}{*}{$\frac{(N_1 -1)(N_1 -2)}{(M_1 -1)(M_1 -2)}$ } &\multirow{2}{*}{$\frac{N_1 -2}{ M_1 -2}$} &$N_1 $& $\frac{N_1 (N_1 -1)}{2}$&$\frac{N_1 -2}{3N_1 }$ &Yes\\
\cline{1-1} \cline{5-8}
\cite{JQ}&
 & & &$\left(\frac{N_1 }{M_1 }\right)^{\frac{(N_1 -1)(N_1 -2)}{(M_1 -1)(M_1 -2)}-1}$& $\frac{N_1 (N_1 -1)}{M_1 (M_1 -1)}$&$<\frac{N_1 -1}{N_1 }\cdot\frac{M_1 -1}{N_1 +M_1 }$ &Yes\\
\hline
\end{tabular}
}
\end{table*}
}

\begin{remark}[Other schemes by using the construction of Theorem \ref{th-design-CDC-rs}]\rm
\label{remark-other}
Similar to the proof of Theorem \ref{th-design-CDC-rs}, we can also obtain a scheme with $r\neq s$  by regarding $\mathfrak{R}$ as the stored file set and   $\mathfrak{A}$ defined in \eqref{eq-function-r-neq-s} as the output function arranged set based on a $t$-$(m,q,M,\lambda)$ GDD. In addition, we can also obtain  two  other  schemes  by regarding the set $\mathfrak{A}$ defined in \eqref{eq-function-r-neq-s} as the stored file set and $\mathfrak{R}$ as the output function arranged set, based on a $t$-$(N,M,\lambda)$ design $(\mathcal{X},\mathfrak{B})$ and $t$-$(m,q,M,\lambda)$ GDD $(\mathcal{X}, \mathfrak{G},\mathfrak{B})$, respectively.  So this implies that our constructing method can adopt other combinatorial structures that have a similar property as the second property of $t$-design.
		\hfill $\square$
\end{remark}

\section{Conclusion}
\label{sec-conclusion}
In this paper, different from the existing constructing methods based on MDS, linear combination, or PDA methods, we constructed two classes of the cascaded CDC schemes with $r=s$ via $t$-design and $t$-GDD. Remarkably, unlike the existing construction methods which require a large operation field and an exponentially large number of input files,  our schemes can notably relax the requirement both on the number of input files and the number of output functions, and simply operate on the binary field while maintaining the one-shot asymptotic optimality. Moreover, for some values of $K$, $r$, and $s$, our schemes achieve smaller
communication load than the state-of-art schemes. Finally, our construction method can be used to construct new schemes for the case $r\neq s$ with small file number and output function number.

\noindent {\bf Acknowledgments.}  The authors  would like to thank the editor and   the anonymous reviewers for their detailed and helpful
 comments and suggestions, which much improved the quality of the paper.

\begin{appendices}

\section{Proof of Lemma \ref{lemm-dual-design-system} and Lemma \ref{lemm-dual-GDD-system}}
\label{appendix-dual-design}
\subsection{Proof of Lemma \ref{lemm-dual-design-system}}
 Assume that $(\mathcal{X},\mathfrak{B})$ is a $t$-$(N,M,\lambda)$ design. Its dual design is  $(\mathcal{V},\mathfrak{R})$. By Definition \ref{def-dual-design}, the dual design has $K=\lambda{N\choose t}/{M\choose t}$ points from \eqref{eq-value-b1} and $N$ blocks.

By the definition of the blocks of $\mathfrak{R}$ in Definition \ref{def-dual-design}, i.e., $\mathcal{B}\in \mathfrak{B}$ is contained by $x\in \mathcal{X} $ if and only if $x\in \mathcal{B}$, we have that
\begin{itemize}
\item the occurrence number of each point of $\mathcal{V}$, say $\mathcal{B}\in\mathcal{V}=\mathfrak{B}$, in $\mathfrak{R}$ equals the size of block $\mathcal{B}\in \mathfrak{B}$. By Definition \ref{def-design}, this number is $M$;
\item the size of each block $\mathcal{R}_x\in \mathfrak{R}$ where $x\in \mathcal{X}$ equals the occurrence number of point $x\in \mathcal{X}$. From \eqref{eq-value-b1}, $\mathcal{R}_x$ has $r=\lambda{N-1\choose t-1}/ \binom{M-1}{t-1}$ points;
\item the cardinality of the intersection of any $t'$ distinct blocks of $\mathfrak{R}$, say $\mathcal{R}_{x_1}$, $\mathcal{R}_{x_2}$, $\ldots$, $\mathcal{R}_{x_{t'}}\in \mathfrak{R}=\mathcal{X}$, equals the number of blocks of $\mathfrak{B}$ containing the subset $\{x_1, x_2, \ldots, x_{t'}\}\subseteq\mathcal{X}$. From \eqref{eq--design-occurrence}, this cardinality is the $
\lambda_{t'}= \lambda{N-t'\choose t-t'}/\binom{M-t'}{t-t'}$.
\end{itemize}
Then the proof is completed.		\hfill $\square$

\subsection{Proof of Lemma \ref{lemm-dual-GDD-system}}
\label{appendix-dual-GDD}

Assume that $(\mathcal{X}, \mathfrak{G}, \mathfrak{B})$ is a $t$-$(m ,q, M,\lambda)$ GDD. Then its dual design is $(\mathcal{V},\mathfrak{R})$. Similar to the proof of Lemma \ref{lemm-dual-design-system}, the first claim can be obtained. In addition, from \eqref{eq-GDD-blocks-number}, the second claim can also be derived.

Now let us consider the third and fourth claims. By Definition \ref{def-dual-design}, when $t'=2$ the cardinality of the intersection of any two distinct blocks, say $\mathcal{R}_{x_1},\mathcal{R}_{x_2}\in \mathfrak{R}$, equals the number of blocks of $\mathfrak{B}$ containing the subset $\{x_1,x_2\}\subseteq\mathcal{X}$. So it is sufficient to consider the occurrence number of any two distinct points $x_1,x_2\in \mathcal{X}$. By Definition \ref{def-GDD} we have the following statements.
\begin{itemize}
\item When points $x_1,x_2$ are in the same group, then there is no block containing the subset $\{x_1,x_2\}$. Clearly there are exactly $m{q\choose 2}$ such distinct points $\{x_1,x_2\}$ since   there are $m$ group and each group has $q$ points;
\item When $x_1,x_2$ are in different groups, from \eqref{eq-GDD-occurrence} there are $\lambda_{2}= \lambda \binom{m-2}{t-2}q^{t-2}/\binom{k-2}{t-2}$ blocks containing $x_1$ and $x_2$. There are ${mq\choose 2}$ point pairs in total, and there are ${m\choose t}{q\choose 2}$ point pairs in the same group. So there are ${mq\choose 2}-{m\choose t}{q\choose 2}$ point pairs distributed in different groups.
\end{itemize} Then the proof is completed.		\hfill $\square$

\section{Proof or Lemma \ref{lem-maximum-gain}} \label{proofOptimal}
For one-shot communications, consider a communication group $\mathcal{S}\subseteq[K]$ where a transmitter $k\in\mathcal{S}$ transmits a one-shot coded signal $X_{k,\mathcal{S}}$ such that all the other workers in $\mathcal{S}\backslash\{k\}$ can decode their desired information. Let $\mathcal{L}\subseteq{S}$ denote the largest leader worker group where every worker in $\mathcal{L}$ requests a distinct intermediate value.

Without loss of generality, let  $\mathcal{S}=\{1,...,|\mathcal{S}|\}$,  $\mathcal{L}=\{1,...,L\}$ with $|\mathcal{L}|=L$, and $X_{k,\mathcal{S}}$ is an arbitrary linear combination of symbols $(V_1,...,V_L)$, e.g., $X_{k,\mathcal{S}}=\oplus_{i=1}^L V_i$, where $V_i$ denotes the symbol associated a file not locally known but desired by the leader worker $i$. Assume each $V_i$ is needed and can be decoded by all workers in a subset $\Gamma_i\subset\mathcal{S}$ based $X_{k,\mathcal{S}}$ and their local intermediate values, then each worker in $\Gamma_i$   must have known all $V_j$, for $j\in\mathcal{L}\backslash\{i\}$. Thus, we have
\begin{IEEEeqnarray*}{rCl}
&&\text{Workers in $\Gamma_1$ know $V_2,V_3,\ldots,V_L$},\\
&&\text{Workers in $\Gamma_2$ know $V_1,V_3,\ldots,V_L$},\\
&&\cdots\\
&&\text{Workers in $\Gamma_L$ know $V_1,V_2,\ldots,V_{L-1}$},
\end{IEEEeqnarray*}
which indicates that each $V_i$ is known by all workers in $\cup_{j\in\mathcal{L}\backslash\{i\}}\Gamma_j$, for all $i\in[L]$.  When   each file is mapped by   $r$ workers and since the worker serving as the transmitter must know all $V_1,V_2,V_3,\ldots,V_L$, we have
\begin{IEEEeqnarray}{rCl}
\cup_{j\in\mathcal{L}\backslash\{i\}} |\Gamma_j| \leq r-1,~\forall i\in[L].
\end{IEEEeqnarray}

Moreover, since each output function is reduced by $s$ workers, we have $|\Gamma_j| \leq s$.

For any one-shot communication group $\mathcal{S}$ described above, each coded signal $X_{k,\mathcal{S}}$ carries useful symbols that are desired and can be decoded by $|\mathcal{S}|-1=  \sum_{i=1}^L |\Gamma_i|$ workers, i.e., the  corresponding multicast gain is $\sum_{i=1}^L |\Gamma_i|$. Thus, the maximum one-shot gain of the communication group $\mathcal{S}$  can be formalized as the following problem
\begin{subequations}\label{ProblemOneshotMaxGain}
\begin{IEEEeqnarray}{rCl}
&&\max_{\Gamma_1,\ldots,\Gamma_L} \sum_{i=1}^L |\Gamma_i|\\
s.t. \ \ &&\bigcup_{j\in\mathcal{L}\backslash\{i\}} |\Gamma_j| \leq r-1\\
&& |\Gamma_j| \leq s.
\end{IEEEeqnarray}
\end{subequations}
From \eqref{ProblemOneshotMaxGain},  we can easily obtain that the    maximum one-shot multicast gain   is   \begin{IEEEeqnarray}{rCl}\label{OneshotMaxGain}
g_\text{max}=sL=r+s-1,%=2\cdot \frac{KM}{N}-1,
\end{IEEEeqnarray}
which is achieved by letting $|\Gamma_j|=s$ and $s(L-1)= r-1$. In particular, when $r=s$, we have $g_\text{max}=2r-1$. Taking the average of one-shot multicast gain over all communication groups, we therefore obtain the upper bound of the maximum one-shot multicast gain  $g^*_\text{one-shot}\leq g_\text{max}$.
		\hfill $\square$
		
\section{Proof of Corollary \ref{orderOptimal}} \label{AppendixAsyOpt}
From \eqref{eq-communication-load} and \eqref{eq-lower-boud} we have \begin{align}\label{eq-raio-max-design}
&\frac{L_{\text{one-shot}}^{*}(r,s)}{L_{t\text{-Design}}}
\leq \frac{\frac{s(1-M/N)}{2r-1}}{\frac{s(1-M/N)}{ 2(r-\lambda_2)}}=\frac{2(r-\lambda_2)}{2r-1}
\\
=&\frac{1-\frac{M-1}{N-1}}
{1-\frac{1}{2\lambda}\cdot\frac{(M-1)!}{(N-1)!}\frac{(N-t)!}{(M-t)!}}
\approx 1\ \  \ (\frac{M}{N}=\frac{r}{K}\rightarrow 0).\nonumber
\end{align}
This implies that our $t$-design scheme in Theorem \ref{th-design-CDC} is asymptotically optimal only if $K\gg r$ for any parameters $t$ and $\lambda$.

By Remark \ref{remark-coded-gain-GDD}, the multicast gain of the scheme in Theorem \ref{th-GDD-CDC} is
$g_{t\text{-GDD}}=2r-\frac{2r(M+q-2)}{mq+q-2}$. From \eqref{eq-communication-load} and \eqref{eq-lower-boud} we have
\begin{align}\label{eq-raio-max-GDD}
\frac{L_{\text{one-shot}}^{*}(r,s)}{L_{t\text{-GDD}}}
&\leq \frac{\frac{s(1-M/N)}{2r-1}}
{\frac{s(1-M/N)}{2r-\frac{2r(M+q-2)}{mq+q-2}}}
=\frac{2r-\frac{2r(M+q-2)}{mq+q-2}}{2r-1}
\\
&=1-\frac{q-1}{ma+q-2}+\frac{mq-M}{mq+q-2}\cdot\frac{1}{2r-1}\nonumber\\
&>1-\frac{q-1}{ma+q-2}+\frac{mq-M}{mq+q-2}\cdot\frac{1}{3r}\nonumber\\
&\approx 1\ \ \ \ \ (\frac{q}{m},\frac{M}{m}\rightarrow 0).\nonumber
\end{align}The last formula holds since
\begin{align*}
&\frac{mq-M}{mq+q-2}\cdot\frac{1}{3r}\\
=&\frac{1}{\lambda}\cdot\frac{mq-M}{mq+q-2}\cdot\frac{M-1}{m-1}
\cdot\frac{M-2}{m-2}\cdots
\cdot\frac{M-t+1}{m-t+1}\\
\approx& 0\ \ \ \ \ (\frac{M}{m}\rightarrow 0).
\end{align*}
This implies that our $t$-GDD scheme in Theorem \ref{th-GDD-CDC}  is asymptotically optimal under one-shot linear delivery if $M/m$ and $q/m$ approximate $0$ for any  parameters $t$ and $\lambda$.	\hfill $\square$

\section{Proof of Corollary \ref{CorollaryThm1vsCDC} }
\label{secConverse-cororllary-2}
By Table \ref{tab-known-CDC}, we see  that our schemes for Theorem \ref{th-design-CDC} and Theorem \ref{th-GDD-CDC} have much smaller $N$ and $Q$ than that of \cite{LMYA}. In the following, we compare their communication loads.

From   \eqref{eq-converse}, the communication load $L_{\text{LMYA}}(r,s)$ of  \cite{LMYA} can be written as
\begin{align*}
&L_{\text{LMYA}}(r,s)=\sum\limits_{l=\max\{r+1,s\}}^{\min\{r+s,K\}}\left( \frac{l-r}{l-1}\cdot \frac{{K-r\choose K-l}{r\choose l-s}}{{K\choose s}}\right)\\
\geq& \frac{s}{r+s-1}\cdot\frac{{K-r\choose K-l}{r\choose r}}{{K\choose s}}\\
=&\frac{s}{r+s-1}\cdot \frac{K-r}{K}\cdot \frac{K-r-1}{K-1}\cdot\frac{K-r-s+1}{K-s+1}\\
\geq& \frac{s}{r+s-1}\cdot \left(1-\frac{r}{K-s+1}\right)^s\\
\approx&\frac{s}{r+s-1} (1+o(1)),\ \ \text{when}\ \  K\gg r,s\ \text{and}\ K\rightarrow \infty.
\end{align*}
When $r=s$,  the above formula can be written as follows.
\begin{align*}
L_{\text{LMYA}}(r,s)> \frac{r}{2r-1} (1+o(1))\ \ \text{if}\    K\gg r\ \text{and}\ K\rightarrow \infty.
\end{align*} Substituting $r=\lambda{N-1\choose t-1}/ \binom{M-1}{t-1}$ into the above formula,  for any positive integers $N$, $m$ and $t$ with $t\leq M<N$ , we have
\begin{align}
L_{\text{LMYA}}(r,s)&>  \frac{\lambda{N-1\choose t-1}/ \binom{M-1}{t-1}}{2\lambda{N-1\choose t-1}/ \binom{M-1}{t-1}-1}\cdot (1+o(1))\ \ \\
&\approx \frac{1}{2},\ \  \text{when}\ \  N\gg M,t\ \text{or}\ N\rightarrow \infty.\label{eqOptAsym}
\end{align}

Now we consider the achievable communication loads in Theorem \ref{th-design-CDC} and Theorem \ref{th-GDD-CDC}, respectively. By Theorem \ref{th-design-CDC}, we have a cascaded CDC scheme with $K=\lambda{N\choose t}/{M\choose t}$ distributed computing workers, $N$ files, $Q$ output functions such that each output function is computed by $s=\lambda{N-1\choose t-1}/ \binom{M-1}{t-1}$ workers, the computation load $r=\lambda{N-1\choose t-1}/ \binom{M-1}{t-1}$, and the communication load $L_{\text{comm.}}=\frac{N-1}{2N}$. Clearly, when $N$ is larger, our communication load $\frac{N-1}{2N}$ approximates $\frac{1}{2}$. Similarly, we can also show that the cascaded CDC scheme in Theorem \ref{th-GDD-CDC} has our communication load approximating $\frac{1}{2}$ when $m$ is large.
	\hfill $\square$

\section{Proof of Corollary \ref{CorollaryThm1vsOthers} and Corollary \ref{remark-performance-GDD-t=2}}
\label{secConverse-cororllary-3-4}
For the $t$-design scheme with $t=2$ in Theorem \ref{th-design-CDC}, denote the corresponding number of files and the number of cached files as $N_1$ and $M_1$, respectively. From Theorem \ref{th-design-CDC}, the communication load of our $t$-design scheme  is $\frac{N_1-1}{2N_1}<\frac{1}{2}$.

Firstly, let us consider the scheme in \cite{WCJ} listed in Table \ref{tab-known-design-CDC-t=2}.  For any positive integers $M_1$ and $N_1$, it is not difficult to check that the following inequality always holds
\begin{align*}
\left(1-\frac{M_1}{N_1}\right)^{\frac{N_1-1}{M_1-1}} \cdot \frac{1}{\frac{2(N_1-1)}{M-1}-1}>  \left(\frac{M_1}{N_1}\right)^{\frac{N_1-1}{M_1-1}}
\end{align*}when
\begin{align}\label{conditionGood}
\frac{ N_1}{M_1}>1+\left(\frac{2 (N_1-1)}{M_1-1}+1\right)^{\frac{M_1-1}{N_1-1}},%\ \ \text{i.e.,}\ \ \frac{K}{r}>1+(2r+1)^{\frac{1}{r}}
\end{align} i.e., $ \frac{K}{r}>1+(2r+1)^{\frac{1}{r}}$,  where $K=\frac{N_1(N_1-1)}{M_1(M_1-1)}$ and $r=\frac{ (N_1-1)}{M_1-1}$, indicating that communication load of scheme in \cite{WCJ} is larger  than 1/2, while ours is smaller than 1/2. Clearly, when $N_1$ is much larger than $M_1$, the above inequality always holds.

Secondly, let us consider the scheme in \cite{JWZ} listed in Table \ref{tab-known-design-CDC-t=2}.  By Definition \ref{def-SBIBD}, we have that the number of workers, the number of files and the number of output functions are equal to $N_1$. Then we have $\frac{N_1(N_1-1)}{M_1(M_1-1)}=N_1
$, i.e., $N_1=M_1^2-M_1+1$. So we have $M_1=\frac{\sqrt{4M_1-3}}{2}$. By Table \ref{tab-known-design-CDC-t=2},  the transmission load in \cite{JWZ} equals
\begin{align*}
&\frac{M_1}{M_1-1}\cdot\frac{N_1-M_1}{N_1}
=\frac{M_1}{M_1-1}\cdot\frac{M_1^2-2M_1+1}{M_1^2-M_1+1}\\
=&1-\frac{1}{M_1^2-M_1+1}\geq \frac{2}{3},\ \ \ \ (M_1\geq 2),
\end{align*} which is larger our communication load $\frac{N_1-1}{2N_1}$.

 Finally, let us consider the scheme in \cite{JQ} listed in Table \ref{tab-known-design-CDC-t=2}. By Table \ref{tab-known-design-CDC-t=2}, the transmission load in   \cite{JQ} equals
\begin{align*}
&\frac{M_1}{M_1-1}\cdot\frac{N_1-M_1}{N_1}
=\left(1+\frac{1}{M_1-1}\right)\cdot\left(1-\frac{M_1}{N_1}\right)
\\
=&1+\frac{N_1-M_1^2}{N_1(M_1-1)}\geq \frac{1}{2},\ \ \ \ (N_1\geq 3M_1, M_1\geq 9),
\end{align*} which is also larger our communication load $\frac{N_1-1}{2N_1}$.
%Since the transmission load of the scheme in \cite{JQ} equals the transmission load of the scheme in \cite{JWZ}, the transmission load is smaller than that of the scheme in \cite{JQ}.

The proof of Corollary \ref{remark-performance-GDD-t=2} is similar to that for Corollary \ref{CorollaryThm1vsOthers}, and is thus omitted.

	\hfill $\square$

\section{Proof of Proposition \ref{pro-required-IVs}}
\label{Appendix-IV-class-1}

When $x=y$, we assume that the intermediate value $v_{x,x}$ is required by a worker $\mathcal{B}\in \mathcal{V}$. From \eqref{eq-function-arrangement-design} we have $ \mathcal{B}\in \mathcal{R}_x$. Clearly worker $\mathcal{B}$ can locally compute it from \eqref{eq-block-design-ZJ}. So we only need to consider the case $x \neq y$. Suppose that all the workers require the intermediate value $v_{y,x}$ and can be locally computed by themselves. From \eqref{eq-function-arrangement-design} and \eqref{eq-block-design-ZJ}, each worker $\mathcal{B}\in \mathcal{R}_y$ then $\mathcal{B}\in \mathcal{R}_x$, i.e., $\mathcal{R}_y\subseteq \mathcal{R}_x$. Since each block of $\mathfrak{R}$ has $r=\lambda{N-1\choose t-1}/ \binom{M-1}{t-1}$ points by the second statement of Lemma \ref{lemm-dual-design-system}. This implies that $|\mathcal{R}_y|=|\mathcal{R}_x|=r$. So we have  $\mathcal{R}_y=\mathcal{R}_x$.
By the third statement of Lemma \ref{lemm-dual-design-system} we have $\lambda_2=|\mathcal{R}_x\bigcap\mathcal{R}_y|=\lambda{N-2\choose t-2}/ \binom{M-2}{t-2}$. Then we have $r=\lambda_2$ which implies $N=M$, a contradiction to Definition \ref{def-design}. Similarly, we can  show the intermediate value $v_{x,y}$ can also be decoded by the desired workers. Then the proof is completed.		\hfill $\square$

\end{appendices}

\bibliographystyle{IEEEtran}
\bibliography{reference}

\end{document}